\documentclass[prd,english,preprintnumbers,amsmath,amssymb,nofootinbib,superscriptaddress,twocolumn]{revtex4-2}

\pdfoutput=1

\usepackage{mathtools}
\usepackage{amsfonts}
\usepackage{mathrsfs}
\usepackage{bbm}
\usepackage{slashed}

\usepackage{graphicx}
\usepackage[dvipsnames]{xcolor}
\usepackage{array}
\usepackage{tikz, pgfplots}
\usepackage[percent]{overpic}
\usetikzlibrary{intersections, pgfplots.fillbetween, patterns, patterns.meta}
\usepgflibrary{patterns}
\usepackage{xspace}
\usepackage{hyperref}

\definecolor{blue}{RGB}{31, 119, 180}
\definecolor{orange}{RGB}{255, 127, 14}
\definecolor{green}{RGB}{44, 160, 44}
\definecolor{red}{RGB}{214, 39, 40}
\definecolor{purple}{RGB}{148, 103, 189}
\definecolor{brown}{RGB}{140, 86, 75}
\definecolor{pink}{RGB}{227, 119, 194}
\definecolor{gray}{RGB}{127, 127, 127}
\definecolor{olive}{RGB}{188, 189, 34}
\definecolor{cyan}{RGB}{23, 190, 207}

\usepackage{xifthen}
\usepackage{dsfont}
\usepackage{appendix}
\usepackage{enumitem}
\usepackage{booktabs}
\usepackage{units}
\usepackage[normalem]{ulem}
\usepackage{soul}

\makeatletter
\g@addto@macro\bfseries{\boldmath}
\makeatother

\setkeys{Gin}{width=0.48\textwidth}

\graphicspath{
	{./}
}

\def\0#1#2{\frac{#1}{#2}}
\def\id{1\!\mbox{l}}

\def\CC{{\mathcal C}}

\newcommand{\Tr}{\mathrm{Tr}}
\newcommand{\STr}{\mathrm{STr}}

\newcommand{\be}{\begin{eqnarray}}
\newcommand{\ee}{\end{eqnarray}}
\newcommand{\del}{\partial}

\newcommand{\nn}{\nonumber }

\newcommand{\beq}{\begin{equation}}
\newcommand{\eeq}{\end{equation}}
\newcommand{\bea}{\begin{eqnarray}}
\newcommand{\eea}{\end{eqnarray}}
\newcommand{\delt}{\del_t}

\newcommand{\Nc}{N_{\rm c}}

\newcommand{\MeV}{\text{MeV}}

\newcommand{\deltapq}[1]{\left(2 \pi\right)^4 \delta^{(4)}\left(#1\right)}

\newcommand{\psib}{\bar{\psi}}

\def\0#1#2{\frac{#1}{#2}}

\newcommand{\orcid}[1]{\href{https://orcid.org/#1}{\includegraphics[height=1.9ex,width=1.9ex]{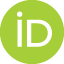}}}

\hypersetup{
	colorlinks,
	linkcolor={red!75!black},
	citecolor={blue!75!black},
	urlcolor={blue!75!black},
	pdftitle={Pressure and speed of sound in two-flavor color-superconducting quark matter at next-to-leading order},
	pdfauthor={Braun, Geissel, Gorda},
	bookmarksopen=true,
	bookmarksopenlevel=2,
	bookmarksnumbered=true
}

\usepackage{babel}
\makeatother
\begin{document}

\title{\texorpdfstring{Pressure and speed of sound in two-flavor color-superconducting quark matter \\
    at next-to-leading order}{Pressure and speed of sound in two-flavor color-superconducting quark matter at next-to-leading order}
	}

\author{Andreas Gei\ss el\,\orcid{0009-0007-9283-4211}\,}
\email{ageissel@theorie.ikp.physik.tu-darmstadt.de}
\affiliation{Institut f\"ur Kernphysik, Technische Universit\"at Darmstadt,
	64289 Darmstadt, Germany}
\author{Tyler Gorda\,\orcid{0000-0003-3469-7574}\,}
\email{gorda@itp.uni-frankfurt.de}
\affiliation{Institut f\"ur Theoretische Physik, Goethe Universit\"at,   Max-von-Laue-Straße 1, 60438 Frankfurt am Main, Germany}
\affiliation{Institut f\"ur Kernphysik, Technische Universit\"at Darmstadt, 64289 Darmstadt, Germany}
\affiliation{ExtreMe Matter Institute EMMI, GSI Helmholtzzentrum f\"ur Schwerionenforschung GmbH, Planckstraße 1, 64291 Darmstadt, Germany}
\author{Jens Braun\,\orcid{0000-0003-4655-9072}\,}
\email{jens.braun@physik.tu-darmstadt.de}
\affiliation{Institut f\"ur Kernphysik, Technische Universit\"at Darmstadt, 64289 Darmstadt, Germany}
\affiliation{ExtreMe Matter Institute EMMI, GSI Helmholtzzentrum f\"ur Schwerionenforschung GmbH, Planckstraße 1, 64291 Darmstadt, Germany}
\begin{abstract}
	Deconfined quark matter at asymptotically high densities is weakly coupled, due to the asymptotic freedom of Quantum Chromodynamics.
	In this weak-coupling regime, bulk thermodynamic properties of quark matter, assuming a trivial ground state, are currently known to partial next-to-next-to-next-to-leading order.
	However, the ground state at high densities is expected to be a color superconductor, in which the excitation spectrum of (at least some) quarks exhibit a gap with a non-perturbative dependence on the strong coupling.
	In this work, we calculate the thermodynamic properties of color-superconducting quark matter at high densities and zero temperature at next-to-leading order (NLO) in the coupling in the presence of a finite gap.
	We work in the limit of two massless quark flavors, which corresponds to deconfined symmetric nuclear matter, and further assume that the gap is small compared to the quark chemical potential.
	In these limits, we find that the NLO corrections to the pressure and speed of sound are comparable in size to the leading-order effects of the gap, and further increase both quantities above their values for non-superconducting quark matter.
	We also provide a parameterization of the NLO speed of sound to guide phenomenology in the high-density region, and we furthermore comment on whether our findings  should be expected to extend to the case of three-flavor quark matter of relevance to neutron stars.
\end{abstract}
\maketitle
%
%
\section{Introduction and Summary}
\label{sec:introsum}
The bulk thermodynamic behavior of strongly interacting matter at large baryon densities and small temperatures remains a fundamental open question within high-energy and nuclear physics.
While at high temperatures and small baryon densities, thermodynamic properties of the fundamental theory of the strong interactions -- Quantum Chromodynamics (QCD) -- can be directly computed with a lattice regularization and Monte-Carlo sampling techniques \cite{Borsanyi:2013bia,HotQCD:2014kol}, such an approach fails in the cold, dense regime of the theory due to the infamous sign problem, see, e.g., Refs.~\cite{deForcrand:2009zkb,Philipsen:2012nu,Aarts:2015tyj,Gattringer:2016kco,Nagata:2021ugx} for reviews.
Despite this difficulty, the study of dense nuclear matter is currently a very active area of research, with interdisciplinary input from astrophysical observations (e.g., Refs.~\cite{Antoniadis:2013pzd,Cromartie:2019kug,Fonseca:2021wxt,TheLIGOScientific:2017qsa,LIGOScientific:2018cki,LIGOScientific:2018hze,Steiner:2017vmg,Nattila:2017wtj,Shawn:2018,Miller:2019cac,Riley:2019yda,Miller:2021qha,Riley:2021pdl}) and theoretical computations in nuclear \cite{Tews:2012fj,Lynn:2015jua,Drischler:2017wtt,Leonhardt:2019fua,Drischler:2020hwi,Keller:2022crb} and high-energy physics \cite{Kurkela:2009gj,Kurkela:2016was,Gorda:2018gpy,Leonhardt:2019fua,Gorda:2021kme,Gorda:2021znl,Gorda:2021gha,Braun:2021uua,Braun:2022jme,Gorda:2023mkk,Gorda:2023usm}.

The input from perturbative computations within QCD itself has recently received increased attention \cite{Komoltsev:2021jzg,Gorda:2022jvk,Somasundaram:2022ztm,Komoltsev:2023zor}, though their inclusion in the determination of the dense-matter equation of state (EOS) within neutron stars was first performed a decade ago \cite{Kurkela:2014vha}.
These computations are convergent above about $25-40$ times nuclear saturation density $n_0 \approx 0.16$~fm$^{-3}$ \cite{Gorda:2023usm} and have currently reached the partial next-to-next-to-next-to-leading order in the strong coupling constant \cite{Gorda:2021kme,Gorda:2021znl,Gorda:2023mkk}.
However, one shortcoming of these computations is that they are performed about the unpaired vacuum, which is unlikely to be the true ground state of QCD at large quark chemical potentials.
Instead, due to the attractive gluonic force between quarks, it is expected that Cooper pairing leads to the generation of a gap in the excitation spectrum of the quarks for at least some species, as demonstrated in various studies based on the fundamental quark and gluon degrees of freedom~\cite{Son:1998uk,Schafer:1999jg,Pisarski:1999bf,Pisarski:1999tv,Brown:1999aq,Evans:1999at,Hong:1999fh,Nickel:2006vf,Braun:2019aow,Leonhardt:2019fua,Braun:2021uua}.
For example, at sufficiently large chemical potential~$\mu$, where the explicit breaking of the flavor symmetry is parametrically suppressed, three-flavor QCD is expected to be a color superconductor in the so-called color-flavor locked (CFL) state where quarks of all three colors and all three flavors form zero-momentum Cooper pairs with zero spin~\cite{Alford:1997zt,Rapp:1997zu,Alford:1998mk,Alford:2002kj}.
This indeed represents the most symmetric way to form Cooper pairs in this regime.
Towards smaller chemical potentials, where the strange quark mass can no longer be neglected, other less symmetric pairing patterns may be more dominant, see Refs.~\cite{Alford:2001dt,Rischke:2003mt,Buballa:2003qv,Shovkovy:2004me,Alford:2007xm,Fukushima:2010bq,Anglani:2013gfu,Baym:2017whm}.
\begin{figure*}[t]
	\includegraphics{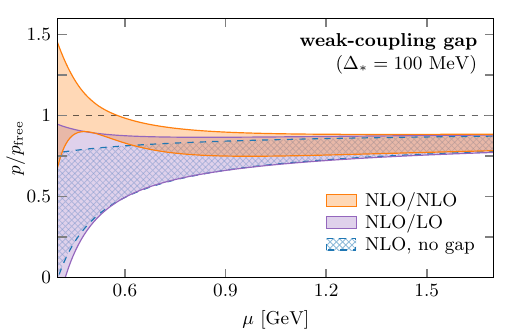}
	\includegraphics{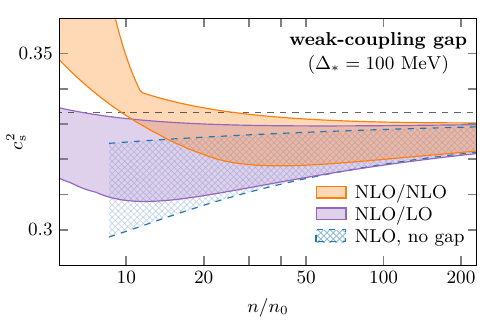}
	\caption{\label{fig:introresult}Pressure normalized to the pressure of the non-interacting quark gas as a function of the quark chemical potential~$\mu$ (left panel) and speed of sound squared as function of the baryon density~$n$ in units of the nuclear saturation density~$n_0$ (right panel) as obtained from a computation at different orders, see main text for details. In our calculations of the pressure and the speed of sound of color-superconducting matter (NLO/LO, NLO/NLO), we have employed the gap found in the weak-coupling limit, see Eq.~\eqref{eq:gapweak}. The results without gap corrections (NLO, no gap) correspond to the well-known perturbative results at two-loop order. The shaded regions depict the uncertainty arising from the usual scale variation of the strong coupling, see Eq.~\eqref{eq:onelooprun}.}
\end{figure*}

The presence of a pairing gap~$|\Delta_\text{gap}|$ in the quark excitation spectrum directly affects thermodynamics properties of strong-interaction matter.
Indeed, the leading-order correction to the pressure of ungapped quark matter is of ${\mathcal O}(\mu^2 |\Delta_\text{gap}|^2)$ (see, e.g., Refs.~\cite{Rajagopal:2000wf,Rajagopal:2000ff,Shovkovy:2002kv,Braun:2021uua,Braun:2022jme}).
Assuming that the QCD ground state is a color superconductor at high densities, this correction must lead to an increase of the pressure relative to the pressure of ungapped quark matter, since the ground state corresponds to the phase with highest pressure.
Depending on the size of the gap and the density under consideration, gap-induced corrections may then already yield a significant contribution to the pressure.
This line of arguments is very general and does not make use of the specific type of the gap.
Moreover, it also applies to situations where the superfluid ground state is associated with the formation of a condensate of color-neutral states, such as pions in the case of large isospin asymmetry at small chemical potential, see, e.g., Refs.~\cite{Brandt:2022hwy,Abbott:2023coj,Fujimoto:2023mvc}.

With respect to phenomenological applications, we add that even if gap-induced corrections to the pressure may be small, the gap may nevertheless strongly affect other observables.
In particular, this applies to quantities which can be derived from the pressure by taking derivatives with respect to the chemical potential -- such as the density and speed of sound -- as well as transport coefficients.
Whether this is the case for specific quantities derived from the pressure or not depends on the scaling of the gap with respect to the chemical potential.
For example, it has been found in a renormalization-group (RG) study based on the fundamental quark and gluon degrees of freedom that the presence of a global maximum in the speed of sound appears tightly connected to the formation of a gap in the quark excitation spectrum~\cite{Leonhardt:2019fua,Braun:2021uua}, at least for two-flavor quark matter.
A more general analysis on how gap-induced corrections to the equation of state affect thermodynamic observables, in particular the speed of sound, can be found in Ref.~\cite{Braun:2022jme}.

In the present work, we consider symmetric nuclear matter in the limit of vanishing temperature and quark masses with a focus on the computation of thermodynamic quantities. To this end, we shall present a framework in Sec.~\ref{sec:theoframework} that allows us to systematically compute the coefficients of an expansion of the pressure~$p$ in powers of the dimensionless gap~$|\bar{\Delta}_\text{gap}|^2 \equiv |{\Delta}_\text{gap}|^2/\mu^2$:
\begin{align}
	p ={}& 	p_{\text{free}}\left ( \gamma_0(g)  + \gamma_1(g) |\bar{\Delta}_{\text{gap}}|^2\right.  \nn\\
	& \qquad\qquad\qquad\quad \left. +\, \gamma_2(g) |\bar{\Delta}_{\text{gap}}|^4 + \dots \right) \,.
	\label{eq:pexp}
\end{align}
Here, $p_{\text{free}}$ is the pressure of the free quark gas, the $\gamma_i$'s denote the expansion coefficients, and~$g$ is the renormalized strong coupling. 
Although we have written the pressure only in terms of monomials of the gap here, it is not excluded that terms with a logarithmic dependence on the gap appear. 
Indeed, starting at~${\mathcal O}(|\Delta_{\text{gap}}|^4)$, also corrections of the form~$\sim |\Delta_{\text{gap}}|^4\ln |\Delta_{\text{gap}}|^2$ can appear as our present study already indicates. 
However, these are irrelevant for our current computation of the coefficients up to~${\mathcal O}(|\Delta_{\text{gap}}|^2)$. 
In any case, we tacitly assume here that the pressure is a smooth function of the gap, which is reasonable away from phase transitions.  
A discussion of other aspects of this expansion of the pressure, such as its form in the presence of finite quark masses can be found in, e.g., Refs.~\cite{Alford:2002kj,Braun:2022jme}.

In our calculations of thermodynamic quantities, we shall restrict ourselves to the case of QCD with two massless quarks, which is simpler than the case with $2+1$ quark flavors from a purely technical standpoint as we do not have to account for the explicit breaking of the flavor symmetry.
In two-flavor quark matter, ignoring the electromagnetic force, the ground state is expected to be governed by pairing of the so-called two-flavor color-superconductor (2SC) type and
the formation of a corresponding chirally symmetric gap in the quark excitation spectrum, as indicated by many non-perturbative first-principles studies~\cite{Son:1998uk,Schafer:1999jg,Pisarski:1999bf,Pisarski:1999tv,Brown:1999aq,Evans:1999at,Hong:1999fh,Nickel:2006vf,Braun:2019aow,Leonhardt:2019fua,Braun:2021uua}.
Note that this type of pairing is different from CFL-type pairing since it does not involve all three colors.

The expansion~\eqref{eq:pexp} of the pressure represents a double expansion in terms of the gap and the strong coupling.
The explicit dependence on the latter is encoded in the coefficients~$\gamma_i$ and arises from quantum corrections to the gapped quark and gluon propagators.
In addition, there is an implicit dependence on the strong coupling coming from the gap.
In contrast to the coefficients~$\gamma_i$, however, the gap is an inherently non-perturbative quantity, representing a non-trivial ground state.
Indeed, color-superconducting gaps are generally found to exhibit a non-analytic dependence on the strong coupling, see, e.g., Refs.~\cite{Son:1998uk,Schafer:1999jg,Pisarski:1999bf,Pisarski:1999tv,Brown:1999aq,Evans:1999at,Hong:1999fh,Nickel:2006vf,Braun:2021uua}.
Hence, a perturbative calculation of the gap in the form of an expansion about~$g=0$ is not possible.
Nevertheless, by treating the strong coupling as a small constant parameter, the dependence of, e.g., the 2SC gap on the strong coupling has been extracted analytically from Dyson-Schwinger equations (DSE)~\cite{Son:1998uk,Schafer:1999jg,Pisarski:1999bf,Pisarski:1999tv,Brown:1999aq,Evans:1999at,Hong:1999fh}:
\begin{align}
	|\Delta_{\text{gap}}| \sim  \mu g^{-5} \exp\left({-\frac{3\pi^2}{\sqrt{2}g}}\right)\,.
	\label{eq:gapweak}
\end{align}
The functional dependence of this gap on the strong coupling is expected to hold for large chemical potentials $\mu\gg \Lambda_{\text{QCD}}$, i.e., at high densities.
Since the strong coupling has been assumed to be small and constant in these studies, the presence of a Cooper instability in finite-density QCD and the associated BCS (Bardeen-Cooper-Schrieffer) mechanism underlie the formation of a gap.
However, towards lower densities, the strong coupling grows and the assumption of a small coupling may no longer be justified.
This increase of the strong coupling may even ``boost'' pair formation which potentially leads to larger gaps and may also alter the dependence of the gap on the strong coupling in Eq.~\eqref{eq:gapweak}, see Refs.~\cite{Braun:2021uua,Braun:2022jme} for a discussion.

In the following we shall not perform non-perturbative computations of the gap.
We rather treat the gap as an input in the presence of which we perform a perturbative computation of the expansion coefficients~$\gamma_i$ in Eq.~\eqref{eq:pexp}.
To be specific, the focus of the present work is on a computation of the pressure, consistently expanded up to second order in the gap and to $\mathcal{O}(g^2)$ in the corresponding coefficients~$\gamma_0$ and~$\gamma_1$.
As we shall discuss in detail in Secs.~\ref{sec:theoframework} and~\ref{sec:pertexp}, we find
\begin{align}
	\gamma_0(g) = 1 - \frac{g^2}{2\pi^2}  + {\mathcal O}(g^4)
	\label{eq:gamma0res}
\end{align}
and
\begin{align}
	\gamma_1(g) = 2 + 1.09(4) g^2 + {\mathcal O}(g^4)\,,
	\label{eq:gamma1res}
\end{align}
the former of which is already well-known~\cite{Baym:1975va,Chapline:1976gq,Freedman:1976xs,Freedman:1976ub,Baluni:1977ms,Toimela:1984xy}.
Thermodynamic stability requires that~$\gamma_1 >0$, provided that the QCD ground state is a color superconductor, but note that the sign of the $\mathcal{O}(g^2)$ corrections in $\gamma_1$ is not similarly fixed.
Note that~$p_{\text{free}}=\mu^4/(2\pi^2)$ in Eq.~\eqref{eq:pexp} for quark matter with two flavors coming in three colors.

The coefficients~$\gamma_0$ and~$\gamma_1$ at order~$g^2$ can be employed to compute thermodynamic quantities, such as the baryon density~$n=(1/3)(\partial p / \partial \mu)$ and the speed of sound~$c_{\rm s}$,
\begin{align}
c_{\rm s} = \left(\frac{\mu}{n} \frac{\partial n}{\partial \mu}\right)^{-1/2} \,,
\label{eq:speed_of_sound}
\end{align}
where $0\leq c_{\rm s}\leq 1$ because of mechanical stability and causality, respectively.
For a (numerical) evaluation of the pressure, which enters these quantities, we also must specify the form of the gap and the strong coupling.
In the following we shall employ the weak-coupling result for the gap, see Eq.~\eqref{eq:gapweak}, with the constant of proportionality adjusted such that~$|\Delta_{\ast}|\equiv|\Delta_{\text{gap}}(n= 10n_0)|=100\,\text{MeV}$.
This represents a typical size of the gap in the literature, see, e.g., Refs.~\cite{Schafer:1999jg,Rischke:2000pv}.
For the strong coupling, we shall for simplicity use the standard one-loop result evaluated at the scale set by the chemical potential:
\begin{align}
	g^2 (2 X\!\mu/\Lambda_{\text{QCD}}) = \frac{1}{b_0\ln (2 X\!\mu/\Lambda_{\text{QCD}})}\,.
	\label{eq:onelooprun}
\end{align}
Here,~$X\in [1/2,2]$ is the usual scale variation factor~\cite{Kurkela:2009gj,Kurkela:2016was,Gorda:2018gpy,Gorda:2021znl,Gorda:2021gha,Gorda:2023mkk,Gorda:2023usm}, which we use to provide  theoretical uncertainty estimates,~$\Lambda_{\text{QCD}} = \Lambda_0 \exp( - 1/( b_0 g^2_0 ))$ with~$b_0 = 29/(24\pi^2)$, and~$g_0$ is the value of the strong coupling at the scale~$\Lambda_0$. In our  numerical calculations, we chose~$g_0^2/(4\pi) \approx 0.179(4)$ and~$\Lambda_0\!=\!10\,\text{GeV}$~\cite{Bethke:2009jm}, which yields~$\Lambda_{\text{QCD}} \approx 265\,\text{MeV}$.

In Fig.~\ref{fig:introresult}, we present our results for the pressure as a function of the quark chemical potential (left panel) and the speed of sound as a function of the baryon density (right panel) as obtained from an evaluation of the coefficients~$\gamma_0$ and~$\gamma_1$ at different orders in~$g$, namely order~$g^0$ (leading order, LO) and order~$g^2$ (next-to-leading order, NLO), respectively.

Examining the pressure in Fig.~\ref{fig:introresult}, we indeed observe that the inclusion of the gap leads to an increase of the pressure relative to the perturbative two-loop result for ungapped quark matter (NLO, no gap), as expected for a color superconductor.
We furthermore see that the NLO corrections to $\gamma_1$ are of a comparable size to (or even larger than) the LO corrections when added on top of the NLO ungapped result and further increase the pressure.
In case of the speed of sound as a function of the baryon density, see the right panel of Fig.~\ref{fig:introresult}, the situation is very similar.
Towards the high-density limit, we observe that the results for the speed of sound with gap corrections approach the value of the free quark gas from below, in accordance with the standard two-loop result.
However, at lower densities, the inclusion of the NLO corrections to $\gamma_1$ computed in this work cause the sound speed to grow even more quickly at lower densities.
They also induce a local minimum in the speed of sound above $30 n_0$, in the perturbative regime.
Decreasing the density further, the gap-induced corrections then even push the speed of sound above its asymptotic high-density value, though in the regime where the strong coupling is no longer small.

Given that the speed of sound in studies based on chiral EFT interactions at low densities has been found to be smaller than~$c_{\rm s}=1/\sqrt{3}$ and decrease towards the zero-density limit (see, e.g., Refs.~\cite{Tews:2012fj,Lynn:2015jua,Drischler:2017wtt,Leonhardt:2019fua}), our present results are in accordance with previous theoretical work~\cite{Leonhardt:2019fua,Braun:2022jme} suggesting a maximum in the speed of sound at high densities.
Such a result also has been found to follow from high-mass neutron-star observations in the three-quark-flavor case, especially once constraints from high-density perturbative-QCD calculations are included in the equation of state inference~\cite{Bedaque:2014sqa,Tews:2018kmu,Greif:2018njt,Annala:2019puf,Huth:2020ozf,Altiparmak:2022bke,Marczenko:2022jhl,Gorda:2022jvk,Annala:2023cwx}.

We emphasize that our general observations regarding the behavior of the pressure and the speed of sound are robust under a variation of the size of the gap at ten times the nuclear saturation density.
With respect to the speed of sound, this essentially only affects the value of the densities at which it exhibits a local minimum or exceeds its asymptotic value.
We discuss this in detail in Sec.~\ref{sec:res} where we vary the size of the gap within a range consistent with recent astrophysical constraints~\cite{Kurkela:2024xfh}.
In that section, we also demonstrate that our observations are additionally robust against a variation of the specific form of the gap used in our numerical computations.
Finally, we present a scaling law for the speed of sound as a function of density in Sec.~\ref{sec:res} which may be useful to guide the construction of parameterizations of the speed of sound in the analysis of astrophysical constraints.

Let us conclude by commenting on how the inclusion of strange quarks may affect our present results.
In three-flavor quark matter, the expansion of the pressure in powers of the gap in Eq.~\eqref{eq:pexp} should assume the same functional form as considered in our present work~\cite{Alford:2002kj,Braun:2022jme}, provided that quark mass corrections are suppressed as they are at large densities \cite{Gorda:2021gha}.
The gap in the expansion would in that case then refer to the CFL gap rather than the 2SC gap.
Since the CFL gap in the weak-coupling limit is also of the form given in Eq.~\eqref{eq:gapweak}, it is reasonable to expect that our conclusions with respect to the pressure and the speed of sound remain unaltered on a qualitative level in the presence of the CFL diquark gap.\footnote{We remark here that in CFL matter there are additional condensates arising from the chiral and baryon symmetry breaking in the CFL phase \cite{Alford:1998mk,Alford:2007xm}, which may complicate the picture. These condensates can however also be handled using the approach employing in the present work, by introducing additional expansion parameters.}
Note that, at the order considered in this work, the coefficient~$\gamma_0^{\text{3-flavor}}$ for three-flavor quark matter agrees identically with the one for two quark flavors~\cite{Baym:1975va,Chapline:1976gq,Freedman:1976xs,Freedman:1976ub,Baluni:1977ms,Toimela:1984xy} as the flavor dependence is absorbed in $p_{\text{free}}^{\text{3-flavor}}=3\mu^4/(4\pi^2)$. The coefficient~$\gamma_1^{\text{3-flavor}}$ is known at ${\mathcal O}(g^0)$ for quark matter in the CFL phase,~$\gamma_1^{\text{3-flavor}}=4$~\cite{Steiner:2002gx,Alford:2002kj,Buballa:2003qv,Alford:2007xm}. Thus, gapped quark matter with two and three flavors behaves qualitatively the same at~NLO/LO.
Of course, a quantitative comparison requires the computation of the corresponding coefficient~$\gamma_1^{\text{3-flavor}}$ at $\mathcal{O}(g^2)$ in $2+1$ flavor QCD, which will be performed in an upcoming work.

\section{Theoretical framework}
\label{sec:theoframework}
In this section, we motivate and describe the computational setup for our double expansion of the pressure in terms of the strong coupling and the gap.
The impatient reader may wish to skip to Subsec.~\ref{subsec:effact} for a concrete discussion of the action that we use.

\subsection{General considerations}
\label{subsec:gencon}
We begin our discussion of the computation of the pressure of dense matter with the action of QCD:
\begin{align}
	S =  \int_0^{\beta}{\rm d}\tau \int {\rm d}^3 x \,  \Big\lbrace & \bar \psi \left(i\slashed{D}- i \mu \gamma_0 \right)\psi \, \nn                    \\
	                                                                & + \frac{1}{4} F_{\mu\nu}^a F_{\mu\nu}^a \Big\rbrace + S_{\rm gf}  + S_{\rm gh} \,.
	\label{eq:QCDaction}
\end{align}
Here, $\beta=1/T$ is the inverse temperature, $\mu$ denotes the quark chemical potential, $F_{\mu\nu}^a$ is the field strength tensor implicitly depending on the non-Abelian gauge field~$A_{\mu}^a$, $S_{\rm gf}$ is the gauge fixing term, and $S_{\rm gh}$ the ghost action. Throughout this work, we consider two massless quark flavors coming in three colors in $3+1$ Euclidean spacetime dimensions.
The coupling between the quark fields~$\psi$ and the gauge fields~$A_{\mu}^a$ (associated with the gluons) is governed by the covariant derivative~$D_\mu$,
\begin{align}
	D_\mu = \del_\mu - i \bar{g} A_\mu^a T^a \,,
\end{align}
where $T^a$ is the generator of the $\text{SU}(3)$ color group in the fundamental representation and~$\bar{g}$ denotes the bare strong coupling.

From the path integral~$\mathcal Z$ as a function of the quark chemical potential,
\begin{align}
	{\mathcal Z}(\mu) = \int {\mathcal D}A_{\mu}^a {\mathcal D}\bar{\psi}{\mathcal D}\psi\, {\rm e}^{-S}\,,
	\label{eq:Zpathint}
\end{align}
the thermodynamic equation of state (i.e., the pressure~$p$) can then in principle be obtained, e.g., at zero temperature:
\begin{align}
	p(\mu) = \lim_{\beta\to\infty} \frac{1}{\beta}\frac{\partial}{\partial V} \ln {\mathcal Z}(\mu)\,,
\end{align}
where $V$ is the spatial volume.

In practice, the computation of the path integral is a highly nontrivial problem and to date requires approximations, in particular at high densities.
The challenges are manifold.
For example, the coupling is weak only at high densities and becomes successively stronger towards the low-density regime.
Moreover, depending on the density, non-perturbative effects leading to a non-analytic dependence on the coupling may become relevant, such as the dynamical formation of a gap in the quark propagator associated with some form of symmetry breaking.
Of course, an analysis of the phenomenological relevance of such non-perturbative effects at high densities requires their inclusion in a computation of the equation of state, which is the main motivation of this work.

Assuming that the coupling is small, the path integral in Eq.~\eqref{eq:Zpathint} may be expanded in terms of it, which yields a corresponding expansion for the pressure and related thermodynamic quantities.
Over many decades now, impressive progress has been made in such computations of thermodynamic properties of cold and dense strong-interaction matter, see, e.g., Refs.~\cite{Freedman:1976xs,Freedman:1976ub,Baluni:1977ms,Toimela:1984xy,Kurkela:2009gj,Gorda:2018gpy,Gorda:2021znl,Gorda:2021kme,Gorda:2023mkk}.
For example, at two-loop order, such a perturbative evaluation of the path integral yields
\begin{align}
	p = p_{\text{free}}\left( 1 - \frac{g^2}{2\pi^2} + {\mathcal O}(g^4) \right)\,.
	\label{eq:p2loop}
\end{align}
Here,~$g$ is the renormalized strong coupling and $p_{\text{free}}$ denotes the pressure of the free/non-interacting quark gas.

Due to the presence of a BCS instability, early ground-breaking non-perturbative studies in the weak-coupling limit pointed out that QCD in the zero-temperature limit at sufficiently high densities is a color superconductor~\cite{Son:1998uk,Schafer:1999jg,Pisarski:1999bf,Pisarski:1999tv,Brown:1999aq,Evans:1999at,Hong:1999fh}.
To be more specific, in the case of QCD with two massless quark flavors with equal chemical potentials, an analysis of the excitation spectrum of the quarks revealed that the formation of a chirally symmetric gap (with quantum numbers $J^{P}=0^{+}$) associated with 2SC pairing is favored at high densities.
Up to a normalization factor, the corresponding gap can in principle be obtained from an evaluation of the following path integral:
\begin{align}
	\Delta_{\text{gap},a} \sim
	\int {\mathcal D}A_{\mu} {\mathcal D}\bar{\psi}{\mathcal D}\psi\, \left(  \bar{\psi}_b  \tau_2 \epsilon_{abc}\gamma_5 {\mathcal C}\bar{\psi}_c^T \right) {\rm e}^{-S}\,,
	\label{eq:gappathint}
\end{align}
where ${\mathcal C}=i\gamma_2\gamma_0$, $\tau_2$ is the second Pauli matrix and the expression is summed over the totally antisymmetric tensor $\epsilon_{abc}$ in color space.
Note that~$|\Delta_{\text{gap}}|^2 = \sum_a \Delta_{\text{gap},a}^{\ast}\Delta_{\text{gap},a}$ is a gauge-invariant quantity.

In order to better understand the role of the gap in two-flavor QCD, it is instructive to look at the path integral from an RG standpoint.
For concreteness, we first consider Wilson's RG approach as it allows us to make direct contact to the path integral in Eq.~\eqref{eq:Zpathint}.
By introducing a momentum scale~$k$ to split the fields in the path integral into soft modes with momenta~$Q^2\lesssim k^2$ and hard modes with~$Q^2\gtrsim k^2$, we can compute the path integral by successively integrating out momentum shells with momenta~$Q^2 \simeq k^2$.
This leads us to a path integral of the following form:
\begin{align}
	{\mathcal Z}(\mu) = \int {\mathcal D}\tilde{A}_{\mu}^a {\mathcal D}\tilde{\bar{\psi}}{\mathcal D}\tilde{\psi}\, {\rm e}^{-S_{\rm W}}\,,
	\label{eq:ZpathintW}
\end{align}
where the path integration is now restricted to soft modes with~$Q^2\lesssim k^2$, and we have added tildes to the fields to denote this change.
The action~$S_{\rm W}$, which is Wilson's effective action, only depends on the soft modes.\footnote{For a discussion of Wilsonian RG flows and gauge invariance, we refer the reader to Ref.~\cite{Litim:1998nf} and references therein.}

Wilson's effective action~$S_{\rm W}$ depends on the so-called RG scale~$k$, $S_{\rm W}\equiv S_{{\rm W},k}$, and its functional form is in general not identical to the original action~$S$.
Instead~$S$ is taken to be the initial point of this so-called RG flow at some high-momentum scale~$k=\Lambda$ in the perturbative regime, $S_{{\rm W},k=\Lambda}=S$, such that also~$\mu/\Lambda \ll 1$.
For~$k<\Lambda$, all terms permitted by the symmetries of the action~$S$ are in general generated, i.e., $S_{\rm W}$ is spanned by an infinite set of scale-dependent couplings.
We emphasize that not all of the generated couplings may be equally relevant for the computation of a given observable.
For example, one-loop corrections to the propagators contribute to the pressure at two-loop order in perturbative calculations of the pressure.
However, one-loop corrections to couplings associated with a four-point function would only contribute to the pressure at three-loop order in perturbative studies.

\begin{figure}[t]
	\vspace{0.14cm}
	\includegraphics[width=0.43\linewidth]{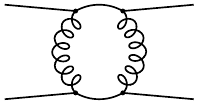}
	$\qquad$
	\includegraphics[width=0.43\linewidth]{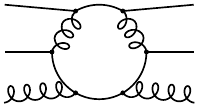}
	\caption{Four-quark interaction (left) and two-gluon-four-quark interactions generating a two-gluon-two-diquark interaction (right). Curly lines represent gluon propagators, while straight lines are fermionic propagators.}
	\label{fig:fund_diags}
\end{figure}

For the generation of a gap in the quark propagator, four-quark interactions~$\lambda_i$ generated via gluon-exchange diagrams in the RG flow are of particular relevance.
Here, the index~$i$ refers to specific symmetry properties of the four-quark channel associated with the coupling~$\lambda_i$.
At one-loop order, we have~$\lambda_i \sim g^4$, i.e., such interactions are dynamically generated by the quark-gluon vertex in~$S$ via two-gluon exchange diagrams as shown in Fig.~\ref{fig:fund_diags}, left; see, e.g., Ref.~\cite{Braun:2011pp} for an introduction.
These interactions directly affect the quark propagator, which can be made explicit by means of a Hubbard-Stratonovich transformation of the path integral.
To be specific, if we are interested in how the aforementioned 2SC-type pairing correlations affect the quark propagator, it is convenient to introduce complex-valued auxiliary (diquark) fields~$\tilde{\Delta}_a$,
\begin{align}
	\tilde{\Delta}_a \sim (\tilde{\bar{\psi}}_b  \tau_2 \epsilon_{abc}\gamma_5 {\mathcal C}\tilde{\bar{\psi}}_c^T)\,,
\end{align}
into the path integral in Eq.~\eqref{eq:ZpathintW},
\begin{align}
	{\mathcal Z}(\mu) = \int  {\mathcal D}\tilde{\Delta}_a^{\ast}  {\mathcal D}\tilde{\Delta}_a{\mathcal D}\tilde{A}_{\mu}^a {\mathcal D}\tilde{\bar{\psi}}{\mathcal D}\tilde{\psi}\, {\rm e}^{-S_{\rm W}}\,,
	\label{eq:ZpathintWHST}
\end{align}
and replace the corresponding dynamically generated four-quark interaction channel associated with the coupling~$\lambda_{2{\rm SC}}$ in $S_{\rm W}$ as follows:
%
	\begin{eqnarray}
	&\lambda_\text{2SC}(\tilde{\psib}_b\tau_2 i\epsilon_{abc}\gamma_5\CC \tilde{\psib}^T_c)(\tilde{\psi}^T_d\CC\gamma_5\tau_2 i \epsilon_{ade}\tilde{\psi}_e)
	\nonumber\\ 
	&\mapsto m^2 \tilde{\Delta}^\ast_a\tilde{\Delta}_a +  i  h\Big[ (\tilde{\psi}^T_b\CC\gamma_5\tau_2\tilde{\Delta}_a \epsilon_{abc}\tilde{\psi}_c)
	\nonumber\\
	&  - (\tilde{\psib}_b\gamma_5\tau_2 \tilde{\Delta}^{\ast}_a \epsilon_{abc}\CC \tilde{\psib}^T_c)\Big]\,,
	\label{eq:hst}
	\end{eqnarray}
%
where~$h^2/m^2=\lambda_\text{2SC}$, see, e.g., Refs.~\cite{Braun:2011pp,Braun:2021uua} for a more detailed discussion.
This transformation may be viewed as a ``coordinate transformation" in the space of operators which span Wilson's effective action~$S_{\rm W}$.
Note that this is an {\it exact} transformation and does not involve an approximation of the original path integral.\footnote{We add that, after having performed a Hubbard-Stratonovich transformation of the path integral, the four-quark interaction channel associated with this transformation will in general be regenerated in the RG flow by the quark-gluon vertex and the Yukawa-like vertex on the right-hand side of Eq.~\eqref{eq:hst}.
Therefore, loosely speaking, Hubbard-Stratonovich transformations have to be performed continuously in the RG flow which can indeed be implemented, see Refs.~\cite{Gies:2001nw,Gies:2002hq,Pawlowski:2005xe,Gies:2006wv,Floerchinger:2009uf,Braun:2014ata,Fu:2019hdw,Fukushima:2021ctq} for discussions of this aspect.}

From the right-hand side of Eq.~\eqref{eq:hst}, we deduce that a nonzero expectation value of the diquark fields generates a gap in the quark propagator,
\begin{align}
	\Delta_{\text{gap},a} \sim \langle \tilde{\bar{\psi}}_b  \tau_2 \epsilon_{abc}\gamma_5 {\mathcal C}\tilde{\bar{\psi}}_c^T\rangle\,,
\end{align}
since the diquark fields couple to two quark fields.
Note that, in our conventions for the Hubbard-Stratonovich transformation, the actual gap in the quark propagator is therefore given by the product of the Yukawa coupling~$h$ and the expectation value of the diquark field.
Since~$\lambda_\text{2SC}\sim g^4$ at one loop order, we have~$h\sim g^2$ for the quark-diquark (Yukawa-type) coupling on the right-hand side of Eq.~\eqref{eq:hst} at this order.\footnote{Here, we exploited the freedom to choose~$h^2=g^4$ in Eq.~\eqref{eq:hst}.
For~$m^2$, we may choose $m^2=\Lambda^2$ where~$\Lambda$ is the scale at which we have performed the Hubbard-Stratonovich transformation.
Other less convenient choices are possible but do not alter our general line of arguments in this work.}	
In practice, it may even be convenient to split the diquark fields into two parts, a potentially spacetime dependent background field and fluctuations about this background:
\begin{align}
	\tilde{\Delta}_a(x_0,\vec{x}^{\,}) \mapsto \tilde{\Delta}_a(x_0,\vec{x}^{\,}) + \delta \tilde{\Delta}_a(x_0,\vec{x}^{\,})\,,
	\label{eq:diquarksplit}
\end{align}
where the background field can be directly related to the gap in the excitation spectrum of the quarks.
Regarding computations of the gap, we note that, from the quark kinetic term in~$S$ and the terms bilinear in the quark fields on the right-hand side of Eq.~\eqref{eq:hst}, we can deduce the parameterization of the quark propagator which has been employed in early non-perturbative Dyson-Schwinger computations of the gap in the weak-coupling limit~\cite{Son:1998uk,Schafer:1999jg,Pisarski:1999bf,Pisarski:1999tv,Brown:1999aq,Evans:1999at,Hong:1999fh}.

The four-quark interactions not only directly affect the quark propagator but also modify the gluon propagator.
This follows immediately from our formulation of Wilson's effective action including the (auxiliary) diquark field $\tilde{\Delta}_a$, which includes quark-diquark interactions; see right-hand side of Eq.~\eqref{eq:hst}.
Indeed, in the RG flow, these quark-diquark interactions generate a covariant term of the form $\sim ( D_{\mu}^{ca} \tilde{\Delta}_a )( D_{\mu}^{cb} \tilde{\Delta}_b)^{\ast}$, which renders the diquark fields dynamical; see, e.g., again Ref.~\cite{Braun:2021uua} for a more detailed discussion.
Note that the diquark fields are not color-neutral fields and therefore naturally couple to the gluons.
By expanding the operator with the covariant derivatives, we obtain, among other terms, a term bilinear in the gauge fields:
\begin{align}
	\sim \tilde{A}_{\mu}^a (T^a)^{bc} \tilde{\Delta}_{c}  (T^e)^{bd} \tilde{\Delta}_{d}^{\ast} \tilde{A}_{\mu}^e\,.
	\label{eq:2g2dq}
\end{align}
This term corresponds to a two-gluon-two-diquark interaction.
Importantly, we observe that a nonzero expectation value of the diquark field not only generates a gap in the quark propagator but potentially also in the gluon propagator.
We add that the interaction in Eq.~\eqref{eq:2g2dq} together with the generation of a finite expectation value of the diquark fields essentially underlies the Anderson-Higgs mechanism~\cite{PhysRev.130.439,Englert:1964et,Higgs:1964ia,Higgs:1964pj,Guralnik:1964eu}, which, in case of pairing of the 2SC type, is associated with the symmetry-breaking pattern~$\text{SU}(3)\to \text{SU}(2)$ in color space.
As a consequence, only three of the eight gluons are massless.
The remaining five gluons are effectively rendered massive by ``eating up'' Goldstone modes that would otherwise appear in the  spectrum associated with the diquark fields, see, e.g., Ref.~\cite{Alford:2007xm} for a review.

We emphasize again that the insertion of the (auxiliary) diquark fields into Wilson's effective action by means of an exact transformation of the path integral is only convenient to make apparent how higher-order quark correlation functions affect the gluon propagator.
In fact, given the definition of the diquark fields, the two-gluon-two-diquark interactions associated with the term in Eq.~\eqref{eq:2g2dq} correspond to two-gluon-four-quark interactions shown in the right panel of Fig.~\ref{fig:fund_diags}.
In a standard loop expansion of the path integral given in Eq.~\eqref{eq:Zpathint}, corrections of this type enter the pressure beginning at four-loop order and are of ${\mathcal O}(g^6)$.

Up to now, we have worked on the level of the path integral and have employed general RG arguments to prepare the discussion of our framework and introduce concepts that underlie our calculation of the QCD pressure in terms of a systematic expansion in powers of the gap.
Standard perturbative-QCD calculations, where the presence of a possible gap is not taken into account, will appear in this expansion at zeroth order in the gap, see Eq.~\eqref{eq:pexp}.
For the actual derivation of this expansion, we shall not operate on the level of the path integral but compute directly the quantum effective action.
To facilitate the systematic setup of the perturbative calculations of the pressure in the presence of a color-superconducting gap we employ as a tool the functional RG (fRG) framework~\cite{Wetterich:1992yh}.
This framework is also a Wilsonian type of RG approach, but operates on the effective action, providing us with direct access to thermodynamic quantities.
The simple one-loop structure of this RG equation (the Wetterich equation) for the effective action makes it advantageous for general discussions and the construction of approximation schemes.

\subsection{Loop expansion of the effective action from the functional Renormalization-Group Approach}
\label{subsec:fRG2loopexpansion}
Thermodynamic quantities can be directly extracted from a computation of the quantum effective action~$\Gamma$.
Indeed, the pressure of a given theory is obtained from an evaluation of the effective action at the ground state~(gs):
\begin{align}
	\label{eq:pressure}
	p = -  \lim_{\beta\to\infty} \frac{1}{\beta}\frac{\partial}{\partial V} \Gamma[\Phi]\Big|_{\text{gs}}\,,
\end{align}
where the vector $\Phi^T=(A^T,\psi^T,\bar{\psi},\dots)$ contains all fields which are employed to span the effective action of a given theory.
The field configuration at the ground state corresponds to a solution of the quantum equation of motion:
\begin{align}
	\left.\frac{\delta\Gamma[\Phi]}{\delta\Phi}\right|_{\text{gs}}=0\,.
\end{align}
As discussed above, in our case of QCD with two massless quark flavors at zero temperature, the field configuration at the ground state is expected to describe a color superconductor of the 2SC type at sufficiently high densities.

In the absence of condensates, the pressure can be obtained from a perturbative expansion of the effective action in the strong coupling~$g$, at least in regimes where the latter is small.
For example, the one-loop effective action is given by
\begin{align}
	\Gamma^{1\mathrm{-loop}}[\Phi] = S[\Phi] + \frac{1}{2}\STr \ln S^{(2)}[\Phi]\Big|_{\text{reg.}} + \text{c.t.}\,,
	\label{eq:Gamma1loop}
\end{align}
where the second term on the right-hand side is assumed to be regularized.
We leave the counter terms (c.t.) unspecified for our general discussion as they depend on the details of the regularization.
The super trace $\STr$ arises since~$\Phi$ contains all types of fields which are employed to span the quantum effective action, and it is defined to provide a minus sign in the fermionic and ghost subspace of the operator~$S^{(2)}$, which is the second functional derivative of the action.
We add that this trace runs over not only discrete indices of~$S^{(2)}$, such as color, flavor, and Dirac indices but also over momenta.

We turn now to the Wetterich equation, which underlies the fRG approach.
It describes the evolution of the scale-dependent effective action~$\Gamma_k$ starting from the classical action~$\Gamma_{k\to\Lambda}=S$ at a suitably chosen initial scale~$\Lambda$ to the full quantum effective action~$\Gamma\equiv\Gamma_{k= 0}$~\cite{Wetterich:1992yh}:
\begin{align}
	\label{eq:Wetterichequation}
	\delt \Gamma_k[\Phi]
	 & = \frac{1}{2} \STr \left[\left(\Gamma^{(2)}_k[\Phi] +R_k\right)^{-1}\cdot\delt R_k\right] \,,
\end{align}
where $t=\ln(k/\Lambda)$ is the so-called RG time.
The regulator function~$R_k$ specifies the Wilsonian momentum-shell integration, such that the RG flow of~$\Gamma_k$ is dominated by fluctuations with momenta~$k$ with~$k^2\sim Q^2$.
In particular, the regulator provides an infrared and ultraviolet regularization of the loop diagrams taking into account in a given study.
According to this equation, the variation of the effective action under the RG scale~$k$ is determined by the full scale-dependent propagator, which is in general a matrix-valued operator also in field space, as it is the case for~$S^{(2)}$:
\begin{align}
	G_k^{-1}=(\Gamma^{(2)}_k+R_k)^{-1}\,.
\end{align}
Looking now at the Wetterich equation from a diagrammatic standpoint, we observe that its right-hand side assumes a simple one-loop structure.
However, this does not imply that only one-loop corrections are included in calculations, as one can in principle systematically generate loop corrections of arbitrarily high orders, which we now illustrate.

The one-loop approximation of the effective action can be immediately seen by replacing~$\Gamma^{(2)}_k$ on the right-hand side of Eq.~\eqref{eq:Wetterichequation} with its initial condition~ $S^{(2)}$ and then integrating over the RG scale~$k$:
\begin{align}
	\Gamma^{1\mathrm{-loop}}_{k}[\Phi] =
	\Gamma_\Lambda^{1\mathrm{-loop}}[\Phi]
	+  \frac{1}{2} \STr   \ln \left( S^{(2)}[\Phi] + R_k \right)  \, ,
	\label{eq:RG1loop}
\end{align}
where
\begin{eqnarray}
	\Gamma_\Lambda^{1\mathrm{-loop}}[\Phi]&=& S[\Phi] + \Gamma_{{\rm c.t.}}^{1\mathrm{-loop}}[\Phi]  \nn\\
	&& \quad\quad -  \frac{1}{2} \STr  \ln \left( S^{(2)}[\Phi] + R_\Lambda\right)\,.
\end{eqnarray}
The term~$\Gamma_{{\rm c.t.}}^{1\mathrm{-loop}}$ depends on the scale~$\Lambda$ and contains the counter terms which are chosen such that $\Gamma^{1\mathrm{-loop}}\equiv \Gamma^{1\mathrm{-loop}}_{k=0}$ does not depend on~$\Lambda$, i.e., it ensures that $\Lambda \partial_{\Lambda}\Gamma^{1\mathrm{-loop}}=0$; see Ref.~\cite{Braun:2018svj} for a detailed discussion.
In particular, this term allows us to remove the scheme dependence introduced by the regulator function.
By comparing Eq.~\eqref{eq:RG1loop} with the standard result for the one-loop effective action in Eq.~\eqref{eq:Gamma1loop}, we can deduce that the regulator renders the loop integrals in a perturbative calculation finite and, by construction, we recover the one-loop effective action in Eq.~\eqref{eq:Gamma1loop} from Eq.~\eqref{eq:RG1loop} in the limit~$k\to 0$ as $\lim_{k\to 0} R_k=0$.

The result for the one-loop effective action in Eq.~\eqref{eq:RG1loop} can then be used to compute the two-loop corrections to the effective action.
More precisely, this is done by now approximating~$\Gamma^{(2)}_k$ in the Wetterich equation~\eqref{eq:Wetterichequation} using the scale-dependent one-loop effective action $\Gamma^{1\mathrm{-loop}}_{k}$ given in Eq.~\eqref{eq:RG1loop}.
After integrating over the RG scale~$k$, this eventually yields the two-loop corrections to the effective action:
\begin{align}
	\Gamma \equiv \Gamma_{k=0} = S + \delta\Gamma_{k=0}^{1\mathrm{-loop}} + \delta\Gamma_{k=0}^{2\mathrm{-loop}}\,,
\end{align}
where, e.g., $\delta\Gamma_{k}^{1\mathrm{-loop}}=\Gamma^{1\mathrm{-loop}}_{k} - S$.
Higher orders in the loop expansion can then be obtained along these lines, see, e.g., Refs.~\cite{Papenbrock:1994kf,Litim:2001ky} for a general discussion.

It may also be worth noting that the Wetterich equation for the effective action can also be used to construct gap equations and study non-perturbative phenomena.
For example, by directly integrating the flow equation~\eqref{eq:Wetterichequation}, we obtain~\cite{Braun:2007bx,Braun:2010cy}:
\begin{align}
	\Gamma [\Phi] = \Gamma_{\Lambda}[\Phi] + \frac{1}{2} \STr \ln \Gamma^{(2)}[\Phi]  +  {\mathcal O}(\partial_t \Gamma_k^{(2)}[\Phi] )
	\label{eq:GammaGap}
\end{align}
with $\Gamma_{\Lambda}[\Phi]$ again containing counter terms.
The third term on the right-hand side corresponds to integrated RG-improvement terms.\footnote{Note that these RG-improvement terms vanish if we identify~$\Gamma_{k}$ with~$S$ on the right-hand side of Eq.~\eqref{eq:GammaGap} and we recover the one-loop effective action in Eq.~\eqref{eq:RG1loop} for $k\to 0$.}
From Eq.~\eqref{eq:GammaGap} we could for example derive the DSE for the quark propagator underlying the computation of the 2SC gap in two-flavor QCD in, e.g., Refs.~\cite{Schafer:1999jg,Pisarski:1999bf,Pisarski:1999tv}.
An explicit derivation and solution of this type of equations is not part of the present work.

From a more general standpoint, our discussion indicates that a solution of the Wetterich equation~\eqref{eq:Wetterichequation} for a given ansatz of the effective action provides an ``interpolation" between regimes which may be well described by a loop expansion of the effective action and regimes which are governed by non-perturbative phenomena.
\subsection{Expansion of the QCD effective action}
\label{subsec:effact}
As our discussion in the previous subsection makes clear, results obtained with non-fRG methods (e.g., the specific coupling dependence of the 2SC gap observed in DSE studies or perturbative results for the QCD equation of state) can also be derived within the fRG approach.
In the following, we shall employ a ``hybrid approach", i.e., we do not solve directly the Wetterich equation~\eqref{eq:Wetterichequation} for a specific ansatz but only employ general RG arguments and properties of the Wetterich equation to underpin the systematic expansion~\eqref{eq:pexp} of the equation of state of QCD at high density in powers of the gap. To this end, parts of our general RG analysis in Subsec.~\ref{subsec:gencon} can indeed be straightforwardly reframed in the fRG approach.

Let us start the RG flow from the QCD action in Eq.~\eqref{eq:QCDaction}, $\Gamma_{\Lambda}=S$.
All operators permitted by the symmetries of QCD will be  generated throughout the RG flow.
As discussed in Sec.~\ref{subsec:gencon}, we can at any RG step perform an exact transformation on the space of operators to introduce complex-valued auxiliary (diquark) fields~$\Delta_a$ which carry the quantum numbers of the operator~$({\bar{\psi}}_b  \tau_2 \epsilon_{abc}\gamma_5 {\mathcal C}{\bar{\psi}}_c^T)$.
After a single infinitesimal RG step, this leads to the following effective action:
\begin{widetext}
	\vspace{-0.3cm}
	\begin{align}
	& \Gamma_{k=\Lambda-\delta k} = \int_0^{\beta}{\rm d}\tau \int {\rm d}^3 x \,\bigg\{ Z_{\psi} \bar{\psi} \left(i\slashed{D}- i \mu \gamma_0 \right)\psi  + \frac{1}{4}Z_A F_{\mu\nu}^{a}F_{\mu\nu}^{a}
	\nn \\
	&
	\qquad\qquad\quad
	+ Z_{\Delta} \left( D_{\mu}^{ca} \Delta_a\right) \left( D_{\mu}^{cb} \Delta_b\right)^{\ast}
	+ 2 \mu  Z_{\Delta} \left(\Delta_a \left(D_{0}^{ab}\Delta_b\right)^{\ast} - \Delta^{\ast}_a \left(D_{0}^{ab}\Delta_b \right)\right)
	 - 4\mu^2  Z_{\Delta} \Delta^\ast_a\Delta_a
	\nn\\
	& \qquad\qquad\quad
	+ \frac{1}{2}  i Z_h {h}(\psi^T_b\CC\gamma_5\tau_2\Delta_a \epsilon_{abc}\psi_c) - \frac{1}{2} i Z_h {h}(\psib_b\gamma_5\tau_2\Delta^\ast_a \epsilon_{abc}\CC \psib^T_c)
	+ Z_{m^2}{m^2} \Delta^\ast_a\Delta_a +\dots
	\bigg\} + S_{\text{gf}} + S_{\text{gh}}
	\,,
	\label{eq:GammaFromS}
	\end{align}
	\vspace{-0.25cm}
\end{widetext}
where $Z_{\psi}$, $Z_A$, $Z_{\Delta}$, $Z_h$, $Z_{m^2}$, $h$, and $m^2$ denote renormalization factors and couplings.
Terms that simultaneously depending on diquark fields and the chemical potential appear since these fields carry a finite baryon number.
The dots on the right-hand side of Eq.~\eqref{eq:GammaFromS} refer to operators of higher orders in the fields and derivatives; see also Ref.~\cite{Reuter:2004kk} for a discussion of the form of the effective action of dense quark matter.

Let us now identify the terms in the effective action Eq.~\eqref{eq:GammaFromS} which are required for a computation of the pressure in Eq.~\eqref{eq:pexp} at a given order in the strong coupling and the 2SC gap.
To this end, it is convenient to split the diquark fields into a constant background field and fluctuations about it, $\Delta_a \to \Delta_a + \delta \Delta_a$, where~$\Delta_a$ now refers to the background. 
Assuming that the 2SC gap is finite, which implies that the background field~$\Delta_a$ is finite, the gluon and quark propagators receive a gap.\footnote{Strictly speaking not all quarks and gluons are gapped in case of a 2SC superconducting ground state since the corresponding condensate only couples two colors. 
See Sec.~\ref{sec:pertexp} below.}
From this and the general form of the flow equation~\eqref{eq:Wetterichequation} for the effective action, it follows that, at one-loop order, the quark propagator already will contribute to {\it all} coefficients of the expansion~\eqref{eq:pexp} of the pressure.
These one-loop contributions do not carry an explicit dependence on the strong coupling~$g$ but only an implicit one via the gap and emerge from an expansion of the quark propagator in powers of the gap, $|\Delta_{\text{gap}}|^2\sim  h^2\sum_a |\Delta_a|^2$.
Note that since the right-hand side of the flow equation~\eqref{eq:Wetterichequation} is determined by the propagator of the fields, one-loop corrections to the quark propagator lead to ${\mathcal O}(g^2)$ contributions to the coefficients~$\gamma_n$, as expected.

The situation is somewhat different for contributions to the pressure from the gluon loop in the flow equation~\eqref{eq:Wetterichequation}.
The appearance of a gap in the gluon propagator arises from a dynamically generated two-gluon-two-diquark coupling in Fig.~\ref{fig:fund_diags} (right), which is included in the term $\left( D_{\mu}^{ca} \Delta_a\right) \left( D_{\mu}^{cb} \Delta_b\right)^{\ast}$.
Because of the six vertices, this term appears at ${\mathcal O}(g^6)$ (see also our discussion in Subsec.~\ref{subsec:gencon}).
Taking into account that it is $|\Delta_{\text{gap}}|^2/h^2$ which contributes to the gluon gap, and furthermore noting that~$h^2 \sim g^4$ at one-loop order (see again Subsec.~\ref{subsec:gencon}), we find that the propagators of the gapped gluons depend on~$g^2 |\Delta_{\text{gap}}|^2$ at leading order with respect to their explicit dependence on the strong coupling.
Moreover, in the medium, the gapped gluon propagators receive corrections from momentum-dependent screening effects of ${\mathcal O}(g^2\mu^2)$ (at one-loop order), see also Refs.~\cite{Rischke:2000qz,Carter:2000gm,Rischke:2001py,Rischke:2002rz}.
From an expansion of the gluon propagators, we therefore deduce that the gluons contribute terms to~$\gamma_0$ which are at least of~${\mathcal O}(g^2)$.
The contributions associated with the gap are irrelevant for this coefficient in the expansion of the pressure.
However, contributions to~$\gamma_1$ requires taking into account both the gap in the gluon propagator and the chemical potential.
Since screening effects are at least of ${\mathcal O}(g^2)$ and the gap enters the propagators in the form of a term $\sim g^2 |\Delta_{\text{gap}}|^2$ at leading order, it follows that contribution to the coefficient~$\gamma_1$ from the gluon loop is at least of~${\mathcal O}(g^4)$.
For the expansion coefficient~$\gamma_2$ associated with terms~$\sim |\Delta_{\text{gap}}|^4$, we also find that the gluon loop in the flow equation of the effective action yields contributions which are at least of~${\mathcal O}(g^4)$.
This reasoning can in principle be continued to coefficients~$\gamma_n$ of higher order.

Our general analysis of the effective action now allows us to systematically construct an action~$S^{\prime}$ which can be employed to compute the coefficients~$\gamma_n$ in the expansion~\eqref{eq:pexp} of the pressure as a function of the strong coupling.
In the present work, we restrict ourselves to the computation of the $g^2$-correction to the coefficient~$\gamma_1$ in Eq.~\eqref{eq:pexp}.
The corresponding correction to the coefficient~$\gamma_0$ is already well-known as it is nothing but the two-loop contribution to the pressure found in conventional perturbative-QCD calculations, which we shall find again from our calculations below.
In any case, from our above discussion of the effective action as generated by the classical QCD action, we deduce that in order to compute the coefficients~$\gamma_0$ and~$\gamma_1$ to $\mathcal{O}(g^2|\Delta_{\rm gap}|^2)$, it suffices to consider the following action~$S^{\prime}$:\footnote{This action is invariant under chiral transformations but not invariant under ${\rm U}(1)_{A}$ transformations.
Note that this is in accordance with an RG analysis of symmetry breaking patterns in dense QCD matter with two quark flavors, where it has been found that a broken~${\rm U}(1)_{A}$ symmetry is required to render 2SC pairing correlations to be the most dominant ones at high density~\cite{Braun:2019aow}, see also Refs.~\cite{Alford:1997zt,Rapp:1997zu,Schafer:1998na,Berges:1998rc} for early model studies.}
\begin{align}
	 S^{\prime} ={}   \int_x& \Big\lbrace \bar{\psi} \left(i\slashed{D}- i \mu \gamma_0 \right)\psi + \frac{1}{4} F_{\mu\nu}^a F_{\mu\nu}^a 
	 \nn \\
	 & + m^2 \Delta_a^*\Delta_a + \frac{1}{2}  i{h}(\psi^T_b\CC\gamma_5\tau_2\Delta_a \epsilon_{abc}\psi_c)  
	 \nn \\
	 & - \frac{1}{2} i {h}(\psib_b\gamma_5\tau_2\Delta^\ast_a \epsilon_{abc}\CC \psib^T_c) \Big\rbrace  + S_\text{gf}+  S_\text{gh}\,.
	 \label{eq:Sprime}
\end{align}
Here, it is in principle not required to consider the diquark field to be constant. However, as we shall only consider constant diquark background fields in our calculations below, we do not include terms with derivatives of the diquark field. 
In particular, as explained above, we do not include terms which give rise to a gap~$\sim g^2 |\Delta_{\text{gap}}|^2$ in the gluon propagators, as it does not contribute to the coefficients~$\gamma_0$ and~$\gamma_1$ at~${\mathcal O}(g^2)$. 
We also note that in order to ensure color neutrality of the system in the 2SC phase, a constant background gauge field $A^8_0$ is also induced, which in turn effectively shifts the chemical potentials of the different quarks by an amount dependent on their color charge~\cite{Gerhold:2003js,Dietrich:2003nu}. 
However, we find that this only induces an $\mathcal{O}(|\Delta_{\rm gap}|^4)$ correction to the pressure, so we do not include this additional background gauge field in our computation.

Before proceeding to the computation of the corrections, let us comment on the parameter~$m^2$ in the action~$S^{\prime}$ and its relation to the 2SC gap.
This parameter together with the quark-diquark interaction~$h$ emerges from the Hubbard-Stratonovich transformation of the four-quark channel associated with 2SC pairing.
From Eq.~\eqref{eq:Sprime}, we now deduce that~$m^2$ can be associated with the curvature of an effective action in terms of the gauge-invariant quantity~$|\Delta|^2$. 
The ground state configuration of the diquark field derived from such an effective action determines the gap in the quark excitation spectrum. 
In the present work, we do not aim at a self-consistent computation of this minimum or the gap, both of which would determine~$m^2$ in terms of the gauge coupling but treat the gap as an external (constant) input parameter for simplicity, $|\Delta_{\text{gap}}| \equiv h |\Delta_{\text{gs}}|$. 
To be more specific, in our computations of the pressure, we shall rather fix $m^2$ to provide a gap that agrees with different previous calculations of the quantity as a function of $\mu$ and $g$. 
In principle, the self-consistently computed solution of the quantum equations of motion of the diquark field is in general not constant. 
Therefore, our assumption of a constant gap is a simplification which may potentially affect our results for~$\gamma_1$ and may introduce a residual gauge dependence in this coefficient. 
We shall at least investigate the gauge dependence further in an upcoming work on three-flavor quark matter.
\section{Perturbative computation of the pressure of gapped quark matter}
\label{sec:pertexp}
The starting point of our perturbative computation of the pressure of gapped two-flavor quark matter up to ${\mathcal O}(g^2)$ in the strong coupling and $|\Delta_\text{gap}|^2$ in the gap is given by the action $S^{\prime}$ in Eq.~\eqref{eq:Sprime}.
For such a perturbative study, it is convenient to introduce Feynman rules as usual.
However, since we have introduced the auxiliary background field~$\Delta_a$ in the action, the quark propagator now assumes the following form:
\begin{align}
	\mathcal{P}_\psi & \equiv
	\begin{pmatrix}
		\langle \psi^T(-P) \psi(Q) \rangle        & \langle \bar\psi(P) \psi(Q) \rangle        \\
		\langle \psi^T(-P) \bar\psi^T(-Q) \rangle & \langle \bar\psi(P) \bar\psi^T(-Q) \rangle \\
	\end{pmatrix}
	\nonumber
	\\
	                 & =
	\begin{pmatrix}
		V_\psi & X_\psi \\
		Y_\psi & W_\psi \\
	\end{pmatrix}
	\deltapq{P-Q} \,
\end{align}
with
\begin{align}
	\label{eq:X}
	X_\psi & \equiv - \Big(\slashed{P}_-G_\psi^- \id_\mathrm{c}
	\nonumber                                                                                                                                                                  \\
	+      & \left[ \slashed{P}_-G_\psi^- - \left(\slashed{P}_- + h^2 |\Delta|^2 \slashed{P}_+ G_\psi^+ \right)G_{\psi,\Delta}^- \right]\epsilon_3^2\Big)\id_\mathrm{f}\,,
	\\
	\label{eq:Y}
	Y_\psi & \equiv - \Big(\slashed{P}_+^T G_\psi^+ \id_\mathrm{c}
	\nonumber                                                                                                                                                                  \\
	+      & \left[ \slashed{P}_+^T G_\psi^+ - \left(\slashed{P}_+ + h^2|\Delta|^2 \slashed{P}_-^T G_\psi^- \right)G_{\psi,\Delta}^+ \right]\epsilon_3^2\Big)\id_\mathrm{f}\,,
	\\
	\label{eq:V}
	V_\psi & \equiv i h \Delta^*  \tau_2 \epsilon_3\left(\slashed P_- \slashed P_+ + h^2 |\Delta|^2 \right)G_{\psi, \Delta}^-G_\psi^+ \gamma_5 \mathcal{C}\,,
\end{align}
and
\begin{align}
	\label{eq:W}
	W_\psi & \equiv - i h \tau_2 \Delta  \mathcal{C} \gamma_5 \left(\slashed P_+ \slashed P_- + h^2 |\Delta|^2 \right)G_{\psi,\Delta}^-G_\psi^+ \epsilon_3^2 \, .
\end{align}
Here, $\id_\mathrm{c}$ and $\id_\mathrm{f}$ are the identity matrices in color and flavor space, respectively, $\tau_2$ is the second Pauli matrix which lives in flavor space, and $(\epsilon_3)_{ab} \equiv \epsilon_{3ab}$ is a totally antisymmetric tensor in color space.
Furthermore, we have introduced $P_\pm \equiv (P_0 \pm i \mu, \vec{p}^{\,})^T$ and the following additional auxiliary quantities:
\begin{align}
	G_\psi^\pm          & \equiv \frac{1}{P_\pm^2},
	\\
	G_{\psi,\Delta}^\pm & \equiv \frac{P_\mp^2}{P_\pm^2P_\mp^2 + 2 h^2 |\Delta|^2 P_\pm \cdot P_\mp+ h^4 |\Delta|^4}\,.
\end{align}
The invariance of the action~$S^{\prime}$ under $\text{SU}(3)$ transformations in color space implies that the effective action and also physical observables only depend on~$|\Delta|^2$. 
In all explicit calculations, we exploit this invariance and have chosen to rotate the background field into the $3$-direction in color space for convenience.

Following our line of arguments in Subsec.~\ref{subsec:effact}, a gap in the gluon propagator leads to corrections to the pressure which are of higher order than those considered in this work.
For the computation of the coefficient~$\gamma_1$ at ${\mathcal O}(g^2)$, it hence suffices to consider the free gluon propagator. In Feynman gauge, which we employ in the present work, we have
\begin{align}
	\left(\mathcal{P}_A^{0}\right)_{\mu\nu}^{ab}
	=
	\frac{1}{P^2} \delta^{ab}
	\delta_{\mu \nu} \deltapq{P-Q} \,,
\end{align}
where $\mu,\nu$ are Lorentz indices and $a,b$ denote adjoint color indices. Moreover, we only need to consider the quark-gluon vertex as derived from the classical action~$S^{\prime}$:
\begin{align}
	(\Gamma^{(3)})_{bc,\mu}^a  =
	\begin{pmatrix}
		0                           & - \bar{g} \left(T_{bc}^a\right)^T \gamma_\mu^T \\
		\bar{g} T_{bc}^a \gamma_\mu & 0                                              \\
	\end{pmatrix}\,.
\end{align}
Here, $\bar{g}$ is the bare strong coupling.

With these prerequisites at hand, we can now compute the effective action as discussed in Subsec.~\ref{subsec:fRG2loopexpansion}.
At one-loop order, we find that only the quarks contribute, since the corresponding gluon and ghost contributions do not depend on the background field and the chemical potential at this order.
To be specific, we find
\begin{align}
	\Gamma^{\rm 1-loop} = & S + \Gamma_\Lambda^{\rm 1-loop} - {\Gamma}^{\rm 1-loop}_\mathrm{quark}
\end{align}
with
\begin{align}
	\label{eq:fermionloopintegral}
	{\Gamma}^{\rm 1-loop}_\mathrm{quark}= 2 V_4  \sum_{\pm} \int_{\vec P}\Bigg(
	 & 2 \sqrt{\left(|\vec P\,|\pm\mu\right)^2+ h^2|\Delta|^2} \nonumber
	\\
	 & +  \sqrt{\left(|\vec P\,|\pm\mu\right)^2} \Bigg)\,.
\end{align}
Here, $\Gamma_\Lambda^{\rm 1-loop}$ contains the counter terms and $V_4$ is the four-dimensional spacetime volume.
We observe that the effective action indeed only depends on the invariant~$|\Delta|^2$, as expected.
For~$|{\Delta}|^2 \to 0$, we recover the pressure of the free quark gas from this expression, i.e., $p_{\rm free} = \mu^4/(2\pi^2)$.
Note that~$p_{\rm free}=-(1/V_4)\Gamma^{\rm 1-loop}|_{|\Delta|^2=0}$ (up to additive constants).

A minimization of the (one-loop) effective potential $U^{1-\text{loop}}=(1/V_4)\Gamma^{1-\text{loop}}$ with respect to $|\Delta|^2$,
\begin{align}
	\left.\frac{\partial U^{1-\text{loop}}}{\partial |\Delta|^2}\right|_{|\Delta_\text{gap}|^2/h^2}=0\,,
\end{align}
determines the parameter $m^2$ in Eq.~\eqref{eq:Sprime} as a function of $|\Delta_\text{gap}|^2$ at this order. Recall that we shall treat the gap as an external parameter to fix~$m^2$ in our present study, as discussed in Subsec.~\ref{subsec:effact}.
In any case, the pressure is then obtained as a function of the dimensionless parameter $|\bar\Delta_\text{gap}|^2 \equiv |\Delta_\text{gap}|^2 / \mu^2$ by evaluating the effective potential at the non-trivial ground state, see Eq.~\eqref{eq:pressure}, i.e., $p=-U(|\Delta_\text{gap}|^2/h^2)$.
Up to order $|\bar\Delta_\text{gap}|^2$, we recover
\begin{align}
	\label{eq:onelooppressure}
	p = p_\text{free}
	\left(1 + 2 |\bar \Delta_\text{gap}|^2 \right) + \mathcal{O}(\bar{\Delta}_\text{gap}^{4})\,,
\end{align}
see, e.g., Refs.~\cite{Rajagopal:2000wf,Rajagopal:2000ff,Alford:2001dt,Shovkovy:2002kv,Rischke:2003mt,Buballa:2003qv,Shovkovy:2004me,Alford:2007xm,Braun:2018svj,Braun:2021uua,Braun:2022olp,Contrera:2022tqh,Ivanytskyi:2022oxv}.
In our terminology, this represents the result for the pressure at LO/LO with~$\gamma_0=1$ and~$\gamma_1=2$ in Eq.~\eqref{eq:pexp}.
\begin{figure*}[t]
	\includegraphics{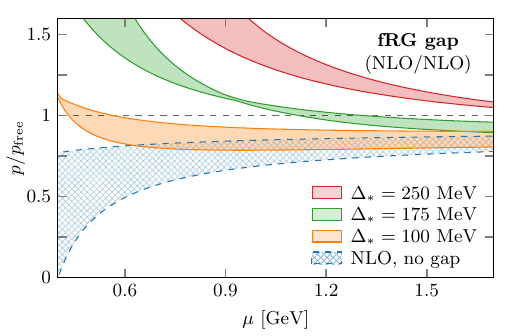}
	\includegraphics{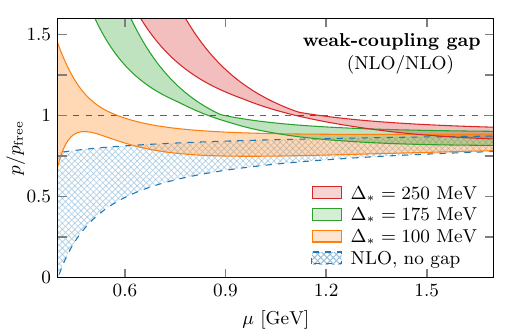}
	\caption{Pressure normalized to the pressure of the non-interacting quark gas as a function of the quark chemical potential~$\mu$. 
		The gap entering the numerical calculations has been adjusted by a constant factor such that it assumes the different values at $n/n_0=10$, $|\Delta_{\ast}|\equiv|\Delta_{\text{gap}}(n = 10 n_0)|$, while leaving their originally predicted dependence on the chemical potential unchanged.
		The corresponding results for ungapped quark matter (NLO, no gap) are shown for comparison.
		Left panel: Pressure as obtained from a computation with the chemical-potential dependence of the gap as found in a recent fRG study \cite{Braun:2021uua}. The uncertainty bands result from the usual scale variation of the strong coupling, see Eq.~\eqref{eq:gapweak}, and the uncertainty band for the fRG gap computed in Ref.~\cite{Braun:2021uua}.
		Right panel: Pressure as obtained from a computation with the chemical potential dependence of the gap as found in the weak-coupling limit, where the uncertainty bands only result from the aforementioned scale variation of the strong coupling.
		\label{fig:EoSp/pfree}
	}
\end{figure*}

To compute the coefficients~$\gamma_0$ and~$\gamma_1$ at ${\mathcal O}(g^2)$, we must consider the following two-loop correction to the effective action:
\begin{align}
	\label{fig:sunset_tikz}
	\begin{aligned}
		\includegraphics[width=0.18\linewidth]{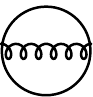}
	\end{aligned}\,\,\,,
\end{align}
where in principle all propagators carry a dependence on the background field.
Note that the effective potential at~$|\bar{\Delta}|^2=0$ at~${\mathcal O}(g^2)$ is directly related to the well-known two-loop result for the pressure of ungapped quark matter, $U(|\bar{\Delta}|^2 = 0)=-p$, which has already been computed analytically many decades ago~\cite{Baym:1975va,Chapline:1976gq,Freedman:1976xs,Freedman:1976ub,Baluni:1977ms,Toimela:1984xy}.
The computation of this diagram including the full dependence on the background field is more complicated.
However, it can be simplified by exploiting the fact that we are only interested in a computation of corrections of~${\mathcal O}(g^2)$ to the pressure up to~${\mathcal O}(|\Delta_{\text{gap}}|^2)$.
For this, it suffices to compute the corrections to the effective potential up to~${\mathcal O}(g^2)$ and~${\mathcal O}(|\Delta|^2)$.
To extract these terms, it is convenient to split the quark propagator into a gapped and ungapped contribution:
\begin{align}
	\label{eq:gapreducedProp}
	\mathcal{P}_{\psi} = \mathcal{P}^0_{\psi} + \left( \mathcal{P}_{\psi} - \mathcal{P}^0_{\psi} \right) \equiv \mathcal{P}^0_{\psi} + \mathcal{P}^\Delta_{\psi}\,.
\end{align}
Here, $\mathcal{P}_{\psi}$ is the quark propagator with the full background-field dependence, ${\mathcal P}^0_{\psi}$ is the propagator without insertions of the background field, and ${\mathcal P}^{\Delta}_{\psi}$ denotes the difference of these two propagators.

We can now expand the above two-loop diagram in terms of~${\mathcal P}^{\Delta}_{\psi}$.
Schematically, this yields
\begin{align}
	\label{eq:2loopdiagamsapprox}
	\hspace{.212cm}
	\begin{aligned}
		\includegraphics[width = .862\linewidth]{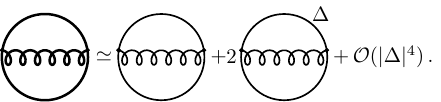}
	\end{aligned}
\end{align}
Here, thick lines represent the propagators with the full background-field dependence, while thin lines correspond to background-field independent propagators and thin lines labeled with $\Delta$ are associated with the ``propagators"~${\mathcal P}^{\Delta}_{\psi}$.
The first diagram on the right-hand side of Eq.~\eqref{eq:2loopdiagamsapprox} does not carry a dependence on the background field and corresponds to the standard two-loop contribution.
From the standpoint of the effective potential, this contribution induces a shift of the potential of ${\mathcal O}(g^2)$, which in turn yields a correction of ${\mathcal O}(g^2)$ to the coefficient~$\gamma_0$ in Eq.~\eqref{eq:pexp}.
The second term on the right-hand side of Eq.~\eqref{eq:2loopdiagamsapprox} is explicitly proportional to $|\Delta|^2$ and generates the correction of ${\mathcal O}(g^2)$ to~$\gamma_1$.

Using the Feynman rules above, we have the following expression for the new term
\begin{align}
	\label{eq:gappedFermionIntegralSchem}
	\Gamma^{2-\text{loop}}_\text{quark}
	={} & \frac{V_4}{2} \int_{P,Q} \left[\mathcal{P}_A^0\right]^{a a'}_{\mu\nu} \left(P-Q\right)
	\Tr \Big[ (\Gamma^{(3)})_{bb',\nu}^{a'}
		\nonumber                                                                                           \\
	    & \times \left[\mathcal{P}_\psi^\Delta\right]_{b' c}\left(P\right) (\Gamma^{(3)})_{cc',\mu}^{a}
		\left[\mathcal{P}_\psi^0\right]_{c' b}\left(Q\right) \Big] \,.
\end{align}
A discussion of the computation of this integral can be found in App.~\ref{app:twoloop}.

From the result of the integral in Eq.~\eqref{eq:gappedFermionIntegralSchem} we can now deduce the correction of $\mathcal{O}(g^2 |\Delta|^2)$ to the effective potential.
Adding this correction and the correction of ${\mathcal O}(g^2)$ from the two-loop integral evaluated with background-field independent propagators to the one-loop effective potential, see Eq.~\eqref{eq:fermionloopintegral}, we obtain the effective potential to the order of interest.
In order to obtain a finite result, we must introduce a counterterm $\delta m^2$ for the diquark mass parameter $m^2$ to cancel an ultraviolet divergence in the integral \eqref{eq:gappedFermionIntegralSchem}.
After renormalization, we can then once again minimize the effective potential and solve for $m^2$, this time to $\mathcal{O}(g^2)$.
Via this procedure, we eventually obtain the pressure up to~${\mathcal O}(|\Delta_{\text{gap}}|^2)$ with coefficients consistently computed up to ${\mathcal O}(g^2)$.
Our results for~$\gamma_0$ and~$\gamma_1$ in the expansion of the pressure in Eq.~\eqref{eq:pexp} can be found in Eqs.~\eqref{eq:gamma0res} and~\eqref{eq:gamma1res}, respectively. Importantly, we note that our results for the expansion coefficients do not depend on a specific form or value of the gap.
\section{Results}
\label{sec:res}
\begin{figure}[t]
	\vspace{0.06cm}
	\includegraphics{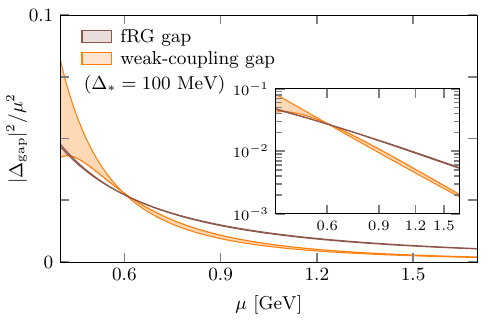}
	\caption{Comparison of the dimensionless gap squared, $|\Delta_{\rm gap}|^2/\mu^2$, as a function of the quark chemical potential, as found in a previous fRG study~\cite{Braun:2021uua} and the weak-coupling limit~\cite{Son:1998uk,Schafer:1999jg,Pisarski:1999bf,Pisarski:1999tv}. 
		The inset shows the same plot on a log-log scale.
		For the comparison, the gaps have been adjusted by a constant factor such that~$\Delta_{\ast}\equiv{|\Delta_{\text{gap}}(n=10n_0)|}=100\,\text{MeV}$.
	}
	\label{fig:gapovermu}
\end{figure}
Let us now use our results for the expansions coefficients~$\gamma_0$ and~$\gamma_1$ in Eq.~\eqref{eq:pexp} to analyze the pressure and speed of sound in the presence of a color-superconducting gap in the quark excitation spectrum.
We compare our results for the pressure and the speed of sound at different orders of the expansion and to perturbative calculations that assume a trivial ground  state.
Furthermore, we study the dependence of these quantities on the size of the gap and on its specific form.
\subsection{Pressure}
In Fig.~\ref{fig:EoSp/pfree} we present our results for the pressure normalized to the pressure of a free non-interacting quark gas as a function of the quark chemical potential for different values~$\Delta_{\ast}$ of the gap at $n=10n_0$, ${\Delta_{\ast} \equiv |\Delta_\mathrm{gap}(n = 10n_0)|}$.
To obtain these results, we have adjusted the gap by rescaling it with a constant factor such that it assumes a given value at $n=10n_0$, while leaving their originally predicted dependence on the chemical potential unchanged.
Note that since the density itself is a derivative of the pressure with respect to the chemical potential, this is an implicit constraint that we solve numerically.
Shown also in this figure is the pressure for unpaired quarks at NLO (labeled ``no gap'').
\begin{figure}[t] 
	\includegraphics{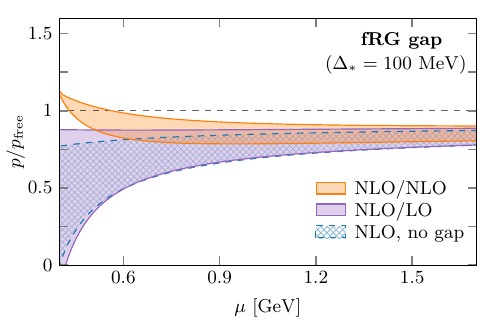}
	\caption{Pressure normalized to the pressure of the non-interacting quark gas as a function of the quark chemical potential~$\mu$ as obtained from a computation at different orders, see main text for details. The gap entering these results has been taken from a previous fRG study~\cite{Braun:2021uua} adjusted by a constant factor such~$\Delta_{\ast}\equiv|\Delta_{\text{gap}}(n=10n_0)|=100\,\text{MeV}$. The results for ungapped quark matter at the corresponding order (NLO, no gap) are shown for comparison. The shaded regions depict the uncertainty arising from the usual scale variation of the strong coupling, see Eq.~\eqref{eq:onelooprun}, and the uncertainty band for the gap given in Ref.~\cite{Braun:2021uua}.}
	\label{fig:EoSp/pfreeLOLOvsNLOLOvsNLONLO}
\end{figure}

We see from the two panels that at the largest chemical potentials -- see especially the right panel of Fig.~\ref{fig:EoSp/pfree} -- all curves approach the unpaired result, as is expected, since gap corrections to bulk thermodynamic properties become smaller at larger chemical potentials as~$|\Delta_{\text{gap}}|/\mu \to 0$.
However, at lower quark chemical potentials the pressure with gap corrections significantly exceeds the pressure of unpaired quarks and, eventually even exceeds the pressure of the non-interacting quark gas (horizontal dashed line).
In all cases, the NLO/NLO pressure exceeds the ungapped pressure, otherwise, the formation of a gap would be disfavored.
Moreover, using continuously smaller and smaller gaps, we eventually smoothly run into the perturbative result.
\begin{figure*}[t] 
	\includegraphics{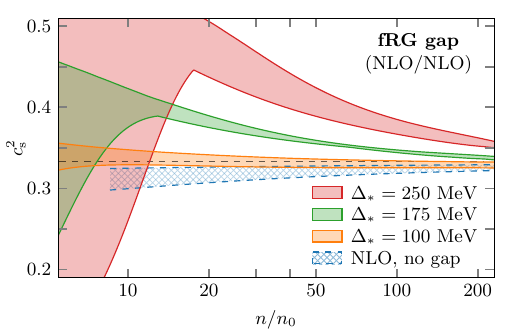}
	\includegraphics{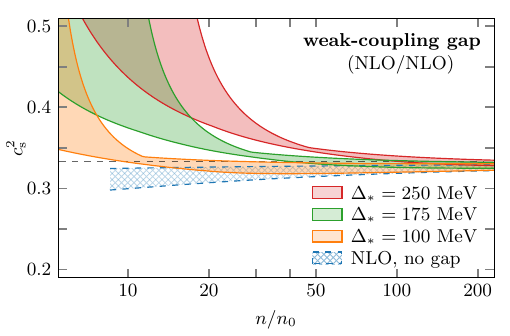}
	\caption{Same as Fig.~\ref{fig:EoSp/pfree}, but for the speed of sound squared $c_{\rm s}^2$ as a function of the number density $n$ in units of the nuclear saturation density $n_0$.}
	\label{fig:SoSdifferentGapSizes}
\end{figure*}

We turn now to a comparison between the two panels of Fig.~\ref{fig:EoSp/pfree}, which illustrate the dependence of the results on the specific functional form of the gap.
The left panel shows our results obtained with the gap from a previous fRG study~\cite{Braun:2021uua}. Although we directly employ the numerical data from the latter work in our numerical calculations, it is instructive to also consider an analytic estimate for the scaling behavior of this gap:
\begin{align}
	\left|\Delta_\mathrm{gap} \right| \sim \Lambda_{\text{QCD}} \exp \left(- \frac{c}{g^4 \mu^2 }\right)\,,
\end{align}
where~$c \sim \Lambda_{\text{QCD}}^2$ is a positive constant. The right panel uses the weak-coupling form of the gap given in Eq.~\eqref{eq:gapweak} above.
Whereas in the computation of the latter, the strong coupling has been assumed to be small and constant, effectively describing the situation at high densities ($\Lambda_{\text{QCD}}/\mu \ll 1$), the former follows from a consideration of lower densities where the density is still high but the coupling can no longer be assumed to be small ($\Lambda_{\text{QCD}}/\mu \lesssim 1$).
Comparing the two panels, we observe that the dependence on the size of the gap is larger in the fRG case at high chemical potentials.
This difference arises because the fRG gap slightly grows as the density is increased within the considered density range, while the weak-coupling result slightly decreases.
However, $|\Delta_\text{gap}|/\mu \to 0$ as $\mu \to \infty$ for both gaps under consideration.
We finally see that for the smallest gap $\Delta_{\ast} = 100\,\mathrm{MeV}$, the results for the two gaps are even quantitatively similar at all densities.
See Fig.~\ref{fig:gapovermu} for a comparison of the two gaps as a function of the chemical potential in this case.

Let us now turn to a comparison of the perturbative result for the pressure at different orders, using the fRG gap, which we show in Fig.~\ref{fig:EoSp/pfreeLOLOvsNLOLOvsNLONLO}.
This serves as a direct comparison with the results shown in Fig.~\ref{fig:introresult} above.
In this figure, we label the different results by a pair of orders, which indicate the order of the expansion used in the $\gamma_0$ and $\gamma_1$ terms, respectively.\footnote{We do not show results at LO/LO. At that order, the coefficient~$\gamma_0$ would be simply a constant which implies that both the pressure normalized to the pressure of the non-interacting quark gas and the speed of sound approach their asymptotic high-density values from above. For completeness, however, we add that the functional form of the expansion~\eqref{eq:pexp} at LO/LO agrees with the one obtained for the pressure in Nambu-Jona-Lasinio-type model calculations at that order, see, e.g.,  Refs.~\cite{Rajagopal:2000wf,Rajagopal:2000ff,Alford:2001dt,Shovkovy:2002kv,Rischke:2003mt,Buballa:2003qv,Shovkovy:2004me,Alford:2007xm,Contrera:2022tqh,Ivanytskyi:2022oxv}.}
In all cases, the overall scale of the gap is fixed to be $\Delta_{\ast} = 100\,\MeV$.
We find that results here are qualitatively similar to the weak-coupling gap shown in Sec.~\ref{sec:introsum}.
In particular, our new NLO/NLO result provides a further small shift to higher pressures over the NLO/LO result.
Though we note that this shift is comparable to, or even larger than, the shift from the LO result.
We see that also for the fRG case, the pressure at some point exceeds the free result, although within the regime of small $\mu$ where the entire perturbative series is not well converged.
\subsection{Speed of Sound}
In this subsection, we analyze the speed of sound $c_\mathrm{s}$, which is more sensitive to corrections arising from the gap because of the additional derivatives with respect to $\mu$ in the definition \eqref{eq:speed_of_sound}.
Additionally, the speed of sound of dense matter is very important from a phenomenological point of view.
It is a direct measure of the stiffness of the equation of state and therefore, e.g., in the three-flavor case at high densities its value is crucial to assessing whether a neutron star of a given mass is stable against gravitational collapse to a black hole.
Due to asymptotic freedom, for high densities (large chemical potentials), the speed of sound approaches the conformal limit of a non-interacting (massless) quark gas of $1/\sqrt{3}$, see the horizontal black dashed line in Fig.~\ref{fig:SoSdifferentGapSizes}.

One recurring question in the very high density regime is whether the speed of sound approaches the conformal limit from above or from below.
Perturbative calculations, which assume a trivial ground state, suggest that the speed of sound approaches the conformal limit from below.
In the current work, we also observe that the speed of sound approaches the limit of a non-interacting quark gas from below, see Fig.~\ref{fig:SoSdifferentGapSizes}, although this may occur only at asymptotically high densities when the gap is large, see in particular the left panel of this figure.

Considering the right panel, we see that the speed of sound falls below the conformal limit starting at the largest densities.
Then, as the density is lowered, it assumes a local minimum and increases, eventually exceeding the conformal limit.
The point at which $c_{\rm s}^2$ crosses~$1/3$ is set by an interplay between the loop corrections in the ungapped result, which drive the speed of sound to smaller values, and corrections from the gap, which drive it to larger ones.
From Fig.~\ref{fig:SoSdifferentGapSizes}, we see that the overall scale of the gap, here set by $\Delta_{\ast}$, is the only relevant parameter that sets where this crossing occurs.
Comparing the fRG and weak-coupling forms of the gap between the two panels, one sees that the same size of the gap at $10n_0$ yields different crossing densities, even though the plot for the pressure, c.f.\ Fig.~\ref{fig:EoSp/pfree}, is quite comparable.
In either case, however, for moderate values of the gap at low densities $\Delta_{\ast} = 100\,\mathrm{MeV}$, the speed of sound remains very close to the conformal value over a wide density range.

Going to lower densities the speed of sound increases for all sizes of the gap $\Delta_{\ast}$, where it must achieve a global maximum before decreasing again to the chiral EFT values at small densities $n \simeq 1-2\,n_0$.
For low densities our approximation breaks down, since the coupling becomes large and the ground state eventually changes from a color superconductor to a state governed by spontaneous chiral symmetry breaking where the BCS gap in the quark spectrum disappears.
An expansion of the pressure in terms of the gap can therefore no longer be considered efficient as the full information on the thermodynamics would then be non-perturbatively encoded in the coefficient~$\gamma_0$.
\begin{figure} 
	\includegraphics{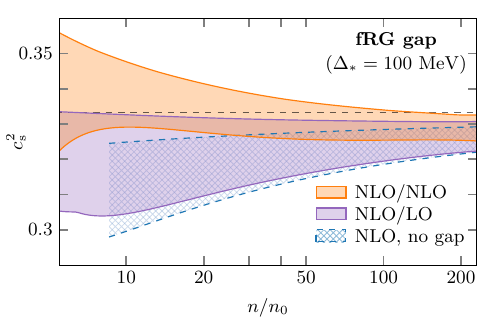}
	\caption{Same as Fig.~\ref{fig:EoSp/pfreeLOLOvsNLOLOvsNLONLO}, but for the speed of sound squared $c_s^2$ as a function of the number density $n$ in units of the nuclear saturation density $n_0$.}
	\label{fig:SoSCombLOvsNLOfRGGap}
\end{figure}

Let us briefly comment on the fRG analogue of the right panel of Fig.~\ref{fig:introresult}, comparing the different approximation schemes already mentioned for the pressure with the fRG form of the gap in Fig.~\ref{fig:SoSCombLOvsNLOfRGGap}.
For high densities, the NLO/LO approximation falls below~$1/3$ and smoothly runs into the NLO (no gap) result.
The next order in our expansion, i.e., NLO/NLO, again pushes the speed of sound slightly upwards.
It however still agrees with the perturbative result at high densities.
We again see broad agreement between the two forms of the gap, with the fRG gap showing slightly larger deviations from the unpaired result at the highest densities.

Up to this point, we have computed the speed of sound by plugging our numerical results for the pressure as a function of the chemical potential into the definition of the speed of sound, see Eq.~\eqref{eq:speed_of_sound}.
Let us finally examine the speed of sound analytically.
To this end, we assume that the gap effectively scales as $|\Delta_{\text{gap}}|\sim \mu^{\sigma}$ where~$\sigma$ is the associated scaling exponent.
As indicated by the inset in Fig.~\ref{fig:gapovermu}, this assumption is indeed reasonable.
Using the numerical data in the range $\mu \in [0.5\,\text{GeV},1.5\,\text{GeV}]$ for $X = 1$, we obtain ${\sigma \approx -0.33}$ for the weak-coupling gap and $\sigma \approx 0.24$ for the fRG gap.
Inspired by the analysis in Ref.~\cite{Braun:2022jme}, we now plug the pressure up to ${\mathcal O}(|\Delta_{\text{gap}}|^2)$ into Eq.~\eqref{eq:speed_of_sound}. This yields
\begin{align}
	c_{\rm s}^2 = \gamma_0^{c_{\rm s}} + \gamma_1^{c_{\rm s}}|\bar{\Delta}_{\text{gap}}|^2 + {\mathcal O}(|\bar{\Delta}_{\text{gap}}|^4)
	\label{eq:csexp}
\end{align}
with
\begin{align}
	\gamma_0^{c_{\rm s}}  = \frac{1}{3} - \frac{b_0}{18\pi^2}g^4 + {\mathcal O}(g^6)\,,
\end{align}
where $b_0$ is defined in Eq.~\eqref{eq:onelooprun} and
\begin{align}
	\gamma_1^{c_{\rm s}} & = \frac{2}{9}(1-\sigma^2)  \nonumber                                                              \\
	                     & \qquad + \frac{1}{9}(1-\sigma^2)\left( \frac{1}{\pi^2} + 1.09(4)\right)g^2 + {\mathcal O}(g^4)\,.
\end{align}
Here, we dropped all terms which are not fully determined by the expansion of the pressure up to ${\mathcal O}(g^2)$ and~${\mathcal O}(|\Delta_{\text{gap}}|^2)$. 
Note that the contribution~$\gamma_0^{c_{\rm s}}$ at ${\mathcal O}(g^4)$ is indeed fully determined by the pressure at~${\mathcal O}(g^2)$.
We add that, for the derivation of the expansion in Eq.~\eqref{eq:csexp}, it is convenient to use~$\mu (\partial g^2/\partial \mu)= -b_0 g^4$, see Eq.~\eqref{eq:onelooprun}, and $\mu(\partial |\bar{\Delta}_{\text{gap}}|^2/\partial\mu) = 2(\sigma-1) |\bar{\Delta}_{\text{gap}}|^2$.

From the expansion in Eq.~\eqref{eq:csexp} we deduce that the speed of sound approaches the value of the conformal limit, $c_{\rm s}^2= 1/3$, from below as~$\sim 1/\ln^2(\mu/\Lambda_{\text{QCD}})$ since the contributions from the gap vanish in the limit~$\mu\to\infty$.
As discussed above, by decreasing the quark chemical potential, the gap then triggers the development of a local minimum in the speed of sound before the associated corrections eventually even push the speed of sound above the value associated with the conformal limit.
This follows from the fact that the coefficient~$\gamma_1^{c_{\text{s}}}$ is strictly positive since we have $\sigma \in (-1,1)$,\footnote{Note that this includes the case of a constant gap,~$\sigma=0$.} at least for the scaling exponent of the gaps considered in our numerical studies.
Put differently, the astrophysically expected existence of a local minimum in the speed of sound at high densities puts a constraint on the scaling of the gap as a function of the chemical potential.
Note also that the increase of the speed of sound towards lower densities is amplified by the correction of~${\mathcal O}(g^2)$ in the coefficient~$\gamma_1^{c_{\rm s}}$.
We expect that these general statements deduced from Eq.~\eqref{eq:csexp} also apply to three-flavor quark matter, provided that quark-mass corrections can be neglected.
Only the numerical values of the expansion coefficients of the speed of sound should be different as the ones for the pressure are in general different for quark matter with two and three flavors, see our discussion in Sec.~\ref{sec:introsum}.

With respect to phenomenological applications, the expansion of the speed of sound given in Eq.~\eqref{eq:csexp} may be considered as a starting point for the construction of parameterizations of the high-density tail of the speed of sound in the analysis of astrophysical constraints.
To this end, the gap could be parameterized by~$|\Delta_{\text{gap}}|=c^{\prime} \mu^{\sigma^{\prime}}$ with $c^{\prime}$ and $\sigma^{\prime}$ being fit parameters.
Of course, this entails that~$\sigma$ in $\gamma_1^{c_{\rm s}}$ should be replaced by the fit parameter~$\sigma^{\prime}$ as well.
For example, performing the fit in this way for our $X = 1$ data as a function of the chemical potential yields $\sigma' \approx -0.35$ for the weak-coupling gap, and $\sigma' \approx 0.24$ for the fRG gap, which agree very well with the direct fits for the scaling exponents, see above.
Exploiting further the relation between the density and the chemical potential for the non-interacting quark gas, the quark chemical potential in the gap and in the coupling can be replaced by~$\mu = c^{\prime\prime}n^{1 / 3}$.
This eventually leads us to a parameterization of the speed of sound as a function of the density which depends on three parameters.
The use of this parameterization or relatives of it in astrophysical applications may provide at least an indirect insight into the properties of color-superconducting matter at high densities.

{\it Acknowledgments.--~}
We thank Aleksi Kurkela, Jan M. Pawlowski, Krishna Rajagopal, and Dirk H. Rischke for helpful discussions.
This work has been supported in part by the Deutsche Forschungsgemeinschaft (DFG, German Research Foundation) project-ID 279384907--SFB 1245 (J.B., A.G., T.G.), by the State of Hesse within the Research Cluster ELEMENTS (projectID 500/10.006) (J.B., T.G.), and by the ERC Advanced Grant ``JETSET: Launching, propagation and emission of relativistic jets from binary mergers and across mass scales'' (Grant No.~884631) (T.G.).
J.B. acknowledges support by the DFG under grant BR~4005/4-1 and~4005/6-1 (Heisenberg program).
%
\appendix
\section{\texorpdfstring{Two-loop Contribution to \\ the Effective Potential}{Two-loop Contribution to the Effective Potential}}
\label{app:twoloop}
In this Appendix, we comment on the calculation of the two-loop effective potential.
By applying the approximations discussed in Sec.~\ref{sec:pertexp}, we obtain the following two-loop integral for the leading background-field dependent contribution on the right-hand side of Eq.~\eqref{eq:2loopdiagamsapprox}:\footnote{The first-term on the right-hand side of Eq.~\eqref{eq:2loopdiagamsapprox} is obtained by replacing $\mathcal{P}_\psi^\Delta$ with $\mathcal{P}_\psi^0$ in Eq.~\eqref{eq:loopintegappedfermion}.}
\begin{widetext}
	\begin{align}
		\frac{\Gamma^{2-\text{loop}}_\text{quark}}{V_4}
		 & = \frac{1}{2} \int_{P,Q} \left[\mathcal{P}_A^0\right]^{a a'}_{\mu\nu} \left(P-Q\right)
		\Tr \Big[ (\Gamma^{(3)})_{bb',\nu}^{a'} \left[\mathcal{P}_\psi^\Delta\right]_{b' c}\left(P\right) (\Gamma^{(3)})_{cc',\mu}^{a}
			\left[\mathcal{P}_\psi^0\right]_{c' b}\left(Q\right) \Big]
		\nonumber
		\\
		 & = g^2 \int_{P,Q} \left[\mathcal{P}_A^0\right]^{a a'}_{\mu\nu} \left(P-Q\right)
		\Tr \big[ \gamma_{\nu} T^{a'}_{bb'}
		\left[X_\psi^\Delta\right]_{b' c}\left(P\right) \gamma_{\mu}T^{a}_{cc'}
		\left[X_\psi^0\right]_{c' b}\left(Q\right) \big] \nonumber
		\\
		 & = 8 g^2  h^2|\Delta|^2  \frac{d_A}{\Nc}  \int_{P,Q} \left[\frac{P_-\cdot Q_-}{P_-^2}\left(h^2|\Delta|^2 + 2 P_+\cdot P_-\right)
			-P_+ \cdot Q_- \right]
		\nonumber                                                                                                                          \\
		 & \hspace{3.8cm}\times\frac{1}{P_+^2P_-^2+h^4|\Delta|^4+2h^2|\Delta|^2 P_+ \cdot P_-}\frac{1}{Q_-^2(P-Q)^2} \, ,
		\label{eq:loopintegappedfermion}
	\end{align}
\end{widetext}
where we have substituted the propagators and the vertices and took the traces in the last step.
Note that~$N_{\rm c}$ is the number of colors and $d_A = N_{\rm c}^2-1$.
We exclusively consider~$N_{\rm c}=3$ in the present work.

Although we are only interested in the computation of two-loop corrections to terms up to $\mathcal{O}(|\Delta|^2)$ in the effective potential, we cannot simply set~$\Delta = 0$ inside the integral on the right-hand side of Eq.~\eqref{eq:loopintegappedfermion} as the background field screens the BCS singularity at $|\vec{p}^{\,}| = \mu$. To resolve this, one may integrate the momenta about the Fermi surface as often done in QCD studies at high densities in the presence of a Cooper instability, see, e.g., Refs.~\cite{Son:1998uk,Schafer:1999jg,Pisarski:1999bf,Pisarski:1999tv,Braun:2021uua}. Either way, the Silver-Blaze symmetry of QCD is broken explicitly, see Refs.~\cite{Khan:2015puu,Braun:2020bhy} for a detailed discussion in this context. 
Here, we choose to leave the background field in the two-loop integral and rearrange the expression by completing squares and compute the integral over $Q$ analytically.
The integral over~$Q$ is then essentially the well-known one-loop quark self-energy in the absence of a background field and requires the standard renormalization to remove an ultraviolet divergence.
Finally, we perform the remaining background-field dependent integral over~$P$ numerically to obtain:
\begin{widetext}
	\begin{align}
		\label{eq:gappedFermionIntegralRes}
		\frac{\Gamma^{2-\text{loop}}_\text{quark}}{V_4}
		 &  =8 g^2  h^2 |\Delta|^2 \mu^2 \frac{d_A }{\Nc} \left(-0.816(1) + 0.0026(1)
		\ln\left(\frac{h^2|\Delta|^2}{\mu^2}\right)\right) + \text{divergences} \, .
	\end{align}
\end{widetext}
To compute the two-loop integral in Eq.~\eqref{eq:loopintegappedfermion}, we have employed a sharp three-momentum cutoff and ensured that our results agree with already existing results, e.g., with the standard one-loop quark self-energy correction in the absence of a background field.
The latter not only underlies the computation of the two-loop pressure of ungapped quark matter but also enters the computation of Eq.~\eqref{eq:gappedFermionIntegralRes}, see also Eq.~\eqref{eq:2loopdiagamsapprox}.
To perform the numerical integration mentioned above, we have first rescaled the dimensionful variables in the integral to be in units of $\mu$.
The numerical integration has been performed for several large values of the dimensionless sharp cutoff $\bar \Lambda \equiv \Lambda/\mu$ and small values of the dimensionless background field~$|\bar \Delta| \equiv |\Delta|/\mu$, i.e., the parameter range for which our approximation is suitable.
The numerical data has then been fitted by positing a suitable fitting function $f(\bar \chi, \bar\Lambda)$, where $\bar \chi \equiv h^2|\bar{\Delta}|^2$.
For vanishing background field, the integral vanishes by construction, which we have also verified numerically.
In particular, this implies that there is no $\bar \Lambda^4$ vacuum contribution needed in the fit function.

\begin{table}[t]
	\begin{tabular}{ l r r r r }
		\toprule
		$c_{i,j}$ & $j=0$        & $j=1$        & $j=2$        & $j=3$                   \\ \midrule
		$i=1$     & $0.0442(0)$  & $0.0017(0)$  & $0.00003(0)$ & $-7.4(0) \cdot 10^{-6}$ \\
		$i=2$     & $-0.816(1)$  & $0.0026(1)$  & $0.450(1)$   & $-0.0017(1)$            \\
		$i=3$     & $-0.0014(1)$ & $-0.0035(1)$ & $0.0015(1)$  & $0.0013(1)$             \\
		\bottomrule
	\end{tabular}
	\caption{\label{tab:ptab}Values of the fit parameters of the ansatz in Eq.~\eqref{eq:fitfunction}.}
\end{table}

To be specific, we have used the following ansatz for $f$, which includes all terms that we would expect to appear as well as terms of one higher order in $\bar{\chi}$ than our results require:
\begin{align}
	f  (\bar \chi, \bar \Lambda)
	= 8 \frac{d_A}{N_c} &\Bigl[  \bar\Lambda^2 \bar\chi \big(c_{1,0}  + c_{1,1} \ln \bar \Lambda
		\nonumber                                                                                                                       \\
	                           & \hspace{0.7cm} +	c_{1,2} \ln \bar\chi + c _{1,3} \ln \bar\chi \ln \bar \Lambda \big)
		\nonumber                                                                                                                       \\
	                           & + \bar\chi (c_{2,0} + c_{2,1} \ln \bar \chi + c_{2,2} \ln \bar \Lambda
		\nonumber                                                                                                                       \\
	                           & \hspace{0.7cm}+ c_{2,3}\ln \bar \chi \ln \bar \Lambda \big)
		\nonumber                                                                                                                       \\
	                           & + \bar \chi^2 \big(c_{3,0} + c_{3,1} \ln \bar \chi
		\nonumber                                                                                                                       \\
	                           & \hspace{0.9cm} + c_{3,2} \ln \bar \Lambda + c_{3,3}\ln \bar \chi \ln \bar \Lambda \big) \Bigr]\, .
	\label{eq:fitfunction}
\end{align}%

\noindent Higher-order terms are not included in our ansatz in Eq.~\eqref{eq:fitfunction}.
The results from this fit can be found in Tab.~\ref{tab:ptab}, where the uncertainties are estimated by additionally performing a few similar fits, using, e.g., functions with some higher-order terms included, or some lower order terms removed to assess the robustness.
To obtain a finite result for the effective potential, we must eventually define a counter term $\delta m^2$ for the diquark mass parameter $m^2$ to cancel the divergent terms in Eq.~\eqref{eq:gappedFermionIntegralRes}, namely all of those that dependent on $\bar{\Lambda}$. 
This renders our results for the effective potential independent of~$\Lambda$.

We close by commenting on the general form of the expansion of the pressure in Eq.~\eqref{eq:pexp}. 
In the two-loop correction to the effective potential, we find terms which depend logarithmically on the gap at ${\mathcal O}( |{\Delta}|^2)$. 
However, after minimizing the (renormalized) effective potential, we find that these terms do not lead to corresponding logarithms in the pressure at $\mathcal{O}(|\Delta|^2)$.
We note however that a $\mathcal{O}(|\Delta|^4 \ln |\Delta|^2)$ term in the effective potential would indeed lead to a $\mathcal{O}(|\Delta|^4 \ln |\Delta|^2)$ term in the pressure.
Hence, we conclude that terms logarithmic in the gap will also appear in the expansion \eqref{eq:pexp} at higher orders.

\bibliography{refs}

\end{document}